\documentclass[aps,pra,showpacs,twocolumn]{revtex4}

\usepackage{amsmath}
\usepackage{amssymb}
\usepackage{mathrsfs}
\usepackage{epsfig}
\usepackage{color}
\usepackage[bookmarks]{hyperref}

\graphicspath{{.}{./EPS/}}

\newcommand* {\bra}[1]{\ensuremath{\langle {#1} |}}
\newcommand* {\ket}[1]{\ensuremath{| {#1} \rangle}}
\newcommand* {\ee}{\ensuremath{\mathrm{e}}}

\newcommand{\cL}{{\mathcal L}}
\newcommand{\cB}{{\mathcal B}}
\newcommand{\cH}{{\mathcal H}}
\newcommand{\cC}{{\mathcal C}}

\newcommand{\cA}{{\mathcal A}}
\newcommand{\cP}{{\mathcal P}}

\newcommand{\cQ}{{\mathcal Q}}
\newcommand{\cO}{{\mathcal O}}
\newcommand{\cD}{{\mathcal D}}
\newcommand{\Tr}[1]{{\textrm{Tr}}{\left\{#1\right\}}}
\newcommand{\TrB}[1]{{\textrm{Tr}_{B}}{\left\{#1\right\}}}
\newcommand{\TrA}[1]{{\textrm{Tr}_{A}}{\left\{#1\right\}}}
\newcommand{\ketbra}[2]{\left|{#1}\right\rangle\left\langle{#2}\right|}

\newcommand{\bet}[1]{|{#1})}
\newcommand{\kra}[1]{({#1}|}
\newcommand{\betkra}[2]{|{#1})({#2}|}
\newcommand{\krabet}[2]{({#1}|{#2})}

\begin{document}
\title{Partly invariant steady state of two interacting open quantum systems}

\author{J\'ozsef Zsolt Bern\'ad}
\email{Zsolt.Bernad@physik.tu-darmstadt.de}
\affiliation{Institut f\"{u}r Angewandte Physik, Technische Universit\"{a}t Darmstadt, D-64289, Germany}
\author{Juan Mauricio Torres}
\affiliation{Institut f\"{u}r Angewandte Physik, Technische Universit\"{a}t Darmstadt, D-64289, Germany}

\date{\today}

\begin{abstract}
We investigate two interacting open quantum systems whose time evolutions are governed by 
Markovian master equations. We show a class of coupled systems whose 
interaction leaves invariant the steady state of one of the systems, i.e., 
only one of the reduced steady states is sensitive to the interactions. 
A detailed proof with the help of the Trotter product formula is presented.
We apply this general statement to a few models, one of which 
is the optomechanical coupling model where 
an optical cavity is coupled to a small mechanical oscillator.
\end{abstract}

\maketitle

\section{Introduction}
Quantum Markovian master equations are the simplest but also the
most used cases of study for the dynamics of open quantum systems.
The concept of open systems plays an important role since perfect isolation 
in the quantum realm and 
access to all degrees of freedom is not possible. 
The main source of inspiration was the
theory of lasers where models describing the interaction of
a central subsystem weakly coupled to an uncontrollable environment were studied.
In the last two decades there has been  great interest in the study of composite quantum systems in order to create entanglement between them \cite{Horodecki}
with the ultimate aim to realize quantum information procedures \cite{Nielsen}. However, 
there is always an environment causing
decoherence which is also the bottleneck to realize highly efficient quantum information protocols. Therefore the theory of open quantum systems was frequently
applied to various models \cite{Aolita}.

There is a subclass of models which can be decomposed into many interacting parts which in turn 
are coupled to independent environments. The best example is the case
of two-level atoms interacting with a single mode of the radiation field and coupled to independent 
reservoirs, like the model of the many-atom laser \cite{Weidlich} or
the one-atom maser \cite{BE}. 
The time evolution of the composite system is derived from a microscopic model or is assembled on the ground of phenomenological considerations from equations which describe independently the parts. 
The latter case usually assumes a low coupling strength between the two parties which allows a separate 
description of dissipative effects.
In both cases the possible treatments of environments consist in Markovian and 
non-Markovian approaches \cite{Beuer}. Obtaining the full solution to the dynamical equation is a very 
complicated task even with a Markovian approach \cite{BE,Torres} and sometimes one has to be satisfied by identifying 
the steady state \cite{Prosen} or its symmetry properties \cite{Albert}.
In general, the role of the interaction in a composed system was studied typically for
two coupled harmonic oscillators \cite{Chimonidou, Dorofeyev} and for two interacting identical atoms \cite{Milonni}. 

In this paper, we focus on the case when two independent quantum systems $A$ and $B$ are subject 
to Markovian master equations and their interaction is studied through
an equation which is obtained by adding the two master equations to a Hamiltonian interaction. 
For physical systems, 
this consideration usually implies that the interaction strength between $A$ and $B$ has to be
the smallest parameter in the system.
We concentrate on a special class of interaction Hamiltonians that cause dephasing on system $A$ 
and lead to a generator (Liouvillian) of the complete system that commutes with the Liouvillian of system $A$'s master equation. 
The motivating example is the optomechanical coupling between 
a single-mode electromagnetic field and a small mechanical oscillator in the presence  
of independent decoherence mechanisms \cite{Vitali}. 
It will be demonstrated that the considered class of interaction Hamiltonians leads to an invariant steady 
state of system $A$, i.e., tracing out system $B$ in the steady state of the composite system one always 
obtains the same steady state of system $A$ which is independent of the interaction Hamiltonian. 
In contrast, the steady state of system $B$ may depend on the interaction.
In the case of finite dimensional systems, theorems of linear algebra  can be employed to prove the statement.
However, when dealing with infinite dimensional systems a different and more sophisticated method is required.
Therefore, the proof of the invariant steady state is based on the properties of quantum dynamical semigroups 
where we employ the Trotter product formula and identify the growth bounds of the master equations. 
While the proof of our statement is rather general we will explicitly show it in some simple models. 
The steady state of system $B$ does not have any particular
property and its solution is model dependent. Therefore in our examples we study the steady state of system $B$
by analyzing its average excitation. Our work presents a general approach  towards composite systems 
governed by independent Markovian master equations. 

This paper is organized  as follows. In Sec. \ref{ex1} we discuss the motivation of our analysis.
In Sec. \ref{ststa} we show the general statement about the steady state. In Sec. \ref{ex}, we apply
these findings to three different systems: two interacting spins, a spin interacting with a harmonic oscillator and two coupled harmonic oscillators.
A discussion about our findings is summarized up in Sec. \ref{Remarks}. In Appendix \ref{Appendix0} we briefly discuss the Liouvillians of the Markovian master equations 
involving some technical problems regarding  infinite dimensional Hilbert spaces. In Appendix \ref{Appendix} we 
analyse the form of the equilibrium state of a damped harmonic oscillator in the number state representation.

\section{Motivation}
\label{ex1}

There are many physical situations in which the dynamics of the system can be decomposed into
two parties subject to a Hamiltonian interaction. The individual sub-systems might 
include a relaxation mechanism that under the Markovian assumption
leads to a master equation for the density matrix
$\hat\varrho$ of the composite system that can be written as
\begin{align}
\frac{d\hat\varrho}{dt}=\cL\hat\varrho=
\mathcal{L}^A\hat\varrho+
\mathcal{L}^B\hat\varrho
+i[\hat\varrho,\hat{H_I}].
 \label{mainEq}
\end{align}
The superoperators $\cL^A$ and $\cL^B$ are generators of the individual systems and $\hat{H}_I$ is 
the interaction Hamiltonian.
One system that allows this type of separation and has gained a lot of recent interest is
the model of optomechanical coupling, where the radiation 
field and a mechanical oscillator are coupled via radiation pressure.
In the single mode scenario, numerical simulations show that the steady state 
of the radiation field remains independent of the coupling strength and of the parameters of the mechanical oscillator. This interesting effect can be generalized
to two systems where one of them is composed of spins and harmonic oscillators and the other system stays arbitrary and the interaction between them holds the same
symmetries as the optomechanical coupling. 
In the following we motivate the problem by showing the optomechanical model.
We consider the simplest Hamiltonian describing the interaction between a single mode of the radiation field and a vibrational mode of a 
mechanical oscillator. This interaction is derived for a cavity with a moving mirror \cite{Law1, Law2} and the quantization of the radiation field is considered
with time-varying boundary conditions. The single mode assumption for both quantized field and mirror motion was successfully applied to describe most of the 
experiments to date and leads to the following interaction Hamiltonian ($\hbar=1$)
\begin{equation}
\hat{H_I}=
g \hat{a}^\dagger \hat{a} (\hat{b}^\dagger+ \hat{b}).
 \label{optoint}
\end{equation}
Thereby, $\hat{a}$ ($\hat{a}^\dagger$) is the annihilation (creation) operator of the single mode 
radiation field with frequency $\omega$ and 
$\hat{b}$ ($\hat{b}^\dagger$) is the annihilation (creation) operator of the moving mirror's mode 
with frequency $\nu$. The parameter $g$ describes the optomechanical coupling strength.
This interaction Hamiltonian commutes with the free Hamiltonian of the radiation field 
$\omega \hat{a}^\dagger \hat{a}$ but it does not commute with the free Hamiltonian of  the mirror 
$\nu \hat{b}^\dagger \hat{b}$. Accounting to an irreversible damping mechanism the Liouville operators of the
individual systems can be cast in the following form
\begin{eqnarray}
\label{optoL}
\mathcal{L}^A\hat\varrho&=&i\omega[\hat\varrho,\hat a^\dagger\hat a]
+\kappa(\bar n+1)\cD[\hat a]\hat\varrho+\kappa\bar n \cD[\hat a^\dagger]\hat\varrho,
\\
\mathcal{L}^B\hat\varrho&=&
i\nu[\hat\varrho,\hat b^\dagger \hat b]+
\gamma(\bar{m}+1)\cD[\hat b]\hat\varrho
+\gamma\bar{m}\cD[\hat b^\dagger]\hat\varrho,
\nonumber 
\end{eqnarray}
which are written in terms of the Lindbladian dissipator 
\begin{equation}
\label{dissipator}
\cD[\hat a]\hat{\varrho}=  \hat a\hat\varrho \hat a^\dagger
-\tfrac{1}{2}(\hat a^\dagger \hat a\hat{\varrho}+\hat\varrho\hat a^\dagger \hat a). 
\end{equation}
The constant $\kappa$ is the loss rate of the radiation field and $\gamma$ 
is the damping rate of the mirror's mode. Both are
assumed to be coupled to a thermal reservoir with 
an average number of photons $\bar{n}$ and phonons $\bar m$.

An interesting feature can be distinguished fairly easily in this model by evaluating
the mean occupation photon and phonon numbers at the steady state of the dynamics.
This calculation is explained in detail in Sec. \ref{ex} and here we merely show the result, namely 
\begin{eqnarray}
\label{meannumbers}
\langle \hat{a}^\dagger \hat{a} \rangle_{\rm st}&=&\bar n, \\
\langle \hat{b}^\dagger \hat{b} \rangle_{\rm st}&=&\bar m+\frac{4 \bar n^2 g ^2}{\gamma^2+4
   \nu^2}+\frac{4 \bar n (\bar n+1)
   (2 \kappa+\gamma) g^2}{\gamma \left[ (2 \kappa+\gamma)^2+4 \nu^2\right]}. \nonumber
\end{eqnarray}
The most striking effect is that the mean photonic occupation number remains unchanged, i.e. it
is insensitive to the coupling to the mechanical oscillator. In contrast the mean phonon number
grows with increasing photon number $\bar n$ and coupling strength $g$. 
In the special case when the photonic bath is at zero temperature ($\bar n=0$),
the occupation numbers of the cavity and the moving mirror remain unchanged. 
It is not a complicated task to show that the composite steady state is insensitive to the interaction
in this situation. This is actually the approximate situation of current  experimental scenarios,
as for optical frequencies $\bar n$ can be regarded to be zero. In addition, the phononic contributions due to the 
coupling in the second line of Eq. \eqref{meannumbers} are negligible in the low
coupling regime  for low values of $\bar n$.
Nevertheless, with the new generations of optomechanical systems reaching the strong coupling regime and for 
microwave frequencies, this effect can become important.
It is worth mentioning that if the coupling is too strong, 
a more detailed model of the loss mechanisms has to be considered such as 
in Ref. \cite{Hu} where the eigenstates of the 
Hamiltonian in Eq. \eqref{optoint} where used to derive the corresponding master equation.  
It can be shown that even in this case the effect persists:
the mean photon number remains unchanged.
We will comment more on this at the end of Sec. \ref{ststa}.

In an attempt to explain the effect one can realize that 
\begin{equation}
 \ee^{-i\hat{H_I}t} \left(\mathcal{L}^A\hat\varrho\right)  \ee^{i\hat{H_I}t}
 =
 \cL^A \left(\ee^{-i\hat{H_I}t} \hat\varrho  \ee^{i\hat{H_I}t}\right).
\end{equation}
It turns out that by identifying this special characteristic one is able to generalize this behaviour.
There are two main purposes of this work. One is to demonstrate that for $\bar n\neq 0$ the steady 
state of the optical mode remains unchanged in contrast to the changing mechanical mode. Second, 
we will extend the result to a broader class of systems that share the same symmetry relations. 
We will show that these arguments
result in an invariant steady state for the system with symmetries and this is 
independent of the details of the other system. 

\section{Steady state analysis}
\label{ststa}

In this section we generalize the statement presented in Sec. \ref{ex1} and prove it by using the properties of dynamical semigroups.  
First, let us consider two quantum systems $A$ and $B$, where each of these systems undergo a Markovian non-unitary dynamics, described by
the Gorini-Kossakowski-Sudarshan-Lindblad generator \cite{Kossakowski,Gorini,Lindblad}. The non-unitary evolution is a trace 
of the interaction of each system 
with an independent environment. These environments are considered to be such that the evolution of systems $A$ and $B$ has a Markovian
irreversible behaviour. The dynamical equation of the joint system 
$d\hat\varrho/dt=\cL\hat\varrho$ is governed by the Liouville operator 
\begin{align}
  \cL\hat\varrho=
  \cL^{A}
  \hat\varrho+
  \cL^{B}\hat\varrho-
  i[\hat V^{A} \hat V^{B},\hat\varrho].
  \label{Liouville}
\end{align}
We assume that system $A$ is composed of $N_P$ harmonic 
oscillators and $N_S$ spins and that its dynamics is 
described by the Lindblad master equation
\begin{align}
  \cL^A\hat\rho^A&=-i[\hat H^A,\hat\rho^A]+\tilde\cL^{A}\hat\rho^A,
  \nonumber\\
  \tilde\cL^A&=
  \sum_{p=1}^{N_P}\cL^{(p)}_{\rm HO}+
  \sum_{s=1}^{N_S}\cL^{(s)}_{\rm S}.
\label{}
\end{align}
The operator $\hat\rho^A$ denotes the reduced density matrix of system $A$ alone and the 
individual Liouvillian for each oscillator has the form
\begin{align}
  \cL^{(p)}_{\rm HO}=
  \kappa_p(\bar n_p+1) \cD[\hat a_p]+ 
  \kappa_p\bar n_p\cD[\hat a_p^\dagger],
  \label{Lharm}
\end{align}
which is written in terms of the dissipator defined in Eq. \eqref{dissipator}.
$\hat a_p$ ($\hat a_p^\dagger$) is the annihilation (creation) operator of the $p$th harmonic oscillator.
Similarly we have for each  spin
\begin{align}
  \cL^{(s)}_{\rm S}=\gamma_s(1-\bar m_s) \cD[\hat\sigma_-^{(s)}]+ 
  \gamma\bar m_s\cD[\hat\sigma_+^{(s)}],
  \label{Lspin}
\end{align}
with the ladder operators $\hat\sigma_{\pm}^{(s)}=(\hat\sigma_{x}^{(s)} \pm i\hat\sigma_{y}^{(s)})/2$ 
expressed
in terms of the Pauli spin operators for the $s$th spin.
All the constants $\gamma_p$ and $\kappa_s$ have dimension of frequency and
are assumed to be non-zero. The dimensionless parameters $\{\bar n_p\geq0\}_{p=1,..,N_p}$ are proportional
to the temperature of the individual thermal reservoir to which each oscillator is coupled and $0\leq\bar m_s\leq1$ for all $s \in \{1,...,N_s\}$.

It is a challenge in the description of a harmonic oscillator that we have to work with unbounded operators, which cannot be defined 
everywhere. This issue is inherited by the Liouvillian in \eqref{Lharm} which is defined on elements of a Banach space. 
In order to discuss the properties of this Liouvillian we require tools from the theory of dynamical semigroups. These tools are supported by Appendix \ref{Appendix0} 
where we show the spectrum of the Liouvillians and we discuss the definition of the adjoint. Examples related to the domain of the harmonic oscillator's
Liouvillian are also presented. 

We assume for system $A$ the following form of the interaction Hamiltonian 
\begin{align}
  \hat V^A=\sum_{p=1}^{N_P}\hat a_p^\dagger \hat a_p +\sum_{s=1}^{N_S} \hat\sigma^+_s\hat\sigma^-_s.
  \label{VA}
\end{align}
The operator $\hat{V}_A$ can be regarded as the excitation operator of system $A$ and therefore it is a dephasing type interaction. Our last assumption is that 
the Hamiltonian of system $A$ commutes with the interaction Hamiltonian,
i.e.
\begin{align}
  [\hat H^A,\hat V^A]=0.
  \label{HA}
\end{align}
Examples of $\hat H^A$ are the Jaynes-Cummings, Ising, and 
Tavis-Cummings Hamiltonians. If these conditions are met, we can state that the steady state of the reduced system 
$A$ is independent of the interaction with system $B$. This can be restated 
as follows: if $\hat\varrho_{\rm st}$ is the steady state of the complete system 
and solution of the equation  $\cL\hat\varrho_{\rm st}=0$, then 
\begin{align}
\cL^A\TrB{\hat\varrho_{\rm st}}=0, 
\label{statement}
\end{align}
and therefore $\hat\rho^A_{\rm st}=\TrA{\hat\varrho_{\rm st}}$ with $\cL^A\hat\rho^A_{\rm st}=0$.
This statement does not hold in general for system $B$ as 
\begin{align}
\cL^B\TrA{\hat\varrho_{\rm st}}\neq 0. 
\label{}
\end{align}

To  prove this statement 
let us start by rewriting the Liouvillian $\cL$ in terms of 
superoperators in the following form
\begin{align}
  \cL=\cL^B+\tilde\cL^A-i\tilde\cC^A-i\cA^A\cC^B-i\cC^A\cA^B
  \label{superL}
\end{align}
where
\begin{align}
  &\cC^i\hat\varrho=[V^i,\hat\varrho],\quad \cA^i\hat\varrho=
  \tfrac{1}{2}\{V^i,\hat\varrho\},\quad
  i\in\{A,B\},
  \nonumber\\
  &\tilde\cC^A\hat\varrho=[H^A,\hat\varrho].
  \label{supers}
\end{align}
One can note that the superoperator $\cC^A$ commutes with the Liouvillian, 
\begin{equation}
\label{LA}
[\cC^A,\cL]=0.  
\end{equation} 
The commutation with the first and last three terms in Eq. \eqref{superL} is evident from Eqs.
\eqref{HA} and \eqref{supers}, and because superoperators acting separately on 
 systems $A$ and $B$ commute. The commutation with $\tilde \cL^A$ can be shown using
the commutation relation between the interaction Hamiltonian $\hat V^A$ 
and the Lindblad operators.

In order to exploit the commutation property between $\cL$ and $\cC^A$, 
it is convenient to work in the eigenbasis of the superoperator $\cC^A$, 
which is defined in \eqref{supers}. As it is formed by the operator $\hat V^A$ 
it is convenient to identify first the eigenvalue equation of $\hat V^A$, i.e. 
\begin{align}
  \hat V^A\ket{n,j}=n \ket{n,j},\quad n,j\in \mathbb{N}.
  \label{VAeig}
\end{align}
Due to \eqref{HA} we have
\begin{equation}
 \hat V^A \hat H^A \ket{n,j}=n \hat H^A \ket{n,j}
\end{equation}
which means that $\hat H^A$ does not couple vectors with different $n$.

The kets $\ket{n,j}$ represent states where the number of photons plus the 
number of excited spins sum to $n$. These states 
have a degeneracy $d_n=d_n(N_P,N_S)$ that is enumerated by the index $j=1,2,\dots d_n$ and depends on the number of spins and 
oscillators in system $A$.
For instance, in the Jaynes-Cummings model these states are
$\ket{n,1}=\ket{n}\otimes\ket{e}$ and $\ket{n,2}=\ket{n+1}\otimes\ket{g}$.
Using the states in \eqref{VAeig} one can  
find that the eigenvectors of the superoperator $\cC^A$ are given by
\begin{align}
  \hat\sigma_{l,\nu}&=
  \ketbra{n+\tfrac{|l|+l}{2},j}{n+\tfrac{|l|-l}{2},k},
  \quad
  \nu=\{n,j,k\},
  \label{}
\end{align}
with $l \in \mathbb{Z}$ and where we have introduced the collective index $\nu$. These eigenvectors are rank $1$ operators and
it can be verified that they solve the eigenvalue equation
\begin{align}
  \cC^A\hat\sigma_{l,\nu}=[\hat V^A,\hat\sigma_{l,\nu}]=l\hat\sigma_{l,\nu}
  \label{eigenC}
\end{align}
with the eigenvalue $l\in \mathbb{Z}$.
As these eigenvectors are degenerate, 
any superposition of $\sigma_{l,\nu}$ with fixed $l$ is also an
eigenvector of $\cC^A$. 
Furthermore, they are orthogonal with respect to  
the Hilbert-Schmidt inner product
\begin{align}
  \krabet{\hat\sigma_{l,\nu}}{\hat\sigma_{l',\nu'}}=
  \TrA{\hat\sigma_{l,\nu}^\dagger\hat\sigma_{l',\nu'}}=
  \delta_{l,l'}\delta_{\nu,\nu'},
  \label{}
\end{align}
where we have introduced for convenience  a ket-bra  notation
$\bet{\hat\sigma_{l,\nu}}$ and $\kra{\hat\sigma_{l,\nu}}$ to write the inner product. 
Let $\cB(\cH)$ be the set of all bounded linear operators on a Hilbert space $\mathcal{H}$.
Given the fact that the states $\ket{n,j}$ are complete in the Hilbert space $\cH^A$ it follows
that the eigenvectors $\hat\sigma_{l,\nu}$ form a complete orthonormal basis 
for the Hilbert space of Hilbert-Schmidt operators 
\begin{equation}
\cB_2(\cH^A):=\left\{\hat x \in \cB(\cH^A):\, \Tr{\hat x^\dagger\hat x}<\infty\right\} 
\end{equation}
which contains the set of all density matrices of system $A$.  
Therefore, in the previously defined ket-bra notation one can  write the  completeness relation as
\begin{align}
  &\sum_{l}\cP_l=\mathbb{I},\quad {\rm with}\quad
  \cP_l=\sum_{\nu}
  \betkra{\hat\sigma_{l,\nu}}{\hat\sigma_{l,\nu}}.
  \label{complete}
\end{align}
Thereby we have introduced the projectors $\cP_l$ that project onto the subspace
of fixed  value of the eigenvalue $l$ of $\cC^A$. 
Using the completeness relation
one can separate the Liouville operator in different subspaces
as $\cL=\sum_{l} \cP_l\cL$ where we have
\begin{equation}
\cP_l\cL=\cL\cP_l. 
\end{equation}
$\cP_l\cL$ acts only onto the subspace
containing eigenelements  $\hat\sigma_{l,\nu}$ of $\cC^A$ for fixed value of $l$.
In particular this superoperator can be written in terms of the projectors as
\begin{align}
  \cP_l\cC^A=\cC^A\cP_l=l\cP_l.
  \label{CPl}
\end{align}
In search of the steady state structure restricted to system $A$ we consider the
reduced dynamical equation and we note that for the subspace labeled by
$l=0$ one has
\begin{align}
  \cP_0\TrB{\cL\hat\varrho}=\left(\tilde\cL^A-i\tilde\cC^A\right)\cP_0\TrB{\hat\varrho}.
  \label{proofl0}
\end{align}
We have used Eqs. \eqref{superL} and \eqref{CPl}, and the relations
$\TrB{\cL^B\hat\varrho}=\TrB{\cC^B\hat\varrho}=0$.
This shows that if $\hat\varrho_{\rm st}$ is the steady state of $\cL$ and
$\hat\rho^A_{\rm st}$ is the steady state of $\cL^A$ then
\begin{align}
  \cP_0\hat\rho^A_{\rm st}=\cP_0\TrB{\hat\varrho_{\rm st}}.
  \label{}
\end{align}
This still does not imply the statement in Eq. \eqref{statement} as we have to show that
the equality also holds for the subspaces $l\neq 0$. 

For the remaining part of the proof we use a more technical approach
based on known results of quantum dynamical semigroups \cite{Ingarden}. 
In this sense, let us recall that the Liouville operator $\cL$ 
in Eq. \eqref{Liouville} is the generator of 
the quantum dynamical semigroup $e^{\cL t}$. 
These two operators act on density matrices $\hat \varrho$ which are trace-class operators in the Hilbert space 
$\cH=\cH^A\otimes\cH^B$. 
The Banach space of all self-adjoint trace-class linear operators containing the set of density matrices can be defined as \cite{Ingarden}
\begin{align}
  L(\cH):=\{\hat \varrho\in \cB(\cH):\, \hat \varrho^\dagger=\hat \varrho, \|\hat  
  \varrho\|_1<\infty\}.
  \label{BanachL1}
\end{align}
with the norm
\begin{align}
  ||\hat \varrho||_1=\Tr{\sqrt{\hat \varrho^\dagger \hat \varrho}}.
  \label{tracenorm}
\end{align}
It was shown by Kossakowski (see Definition $1$ and Theorem $1$ in \cite{Ingarden}) 
that the Liouville operator $\cL$ generates a contraction semigroup on $L(\cH)$. This means that 
the norm of the evolution operator is bounded as 
$\|e^{\cL t}\|_{\rm op} \le 1$ which implies that the norm 
of any evolved element $e^{\cL t}\hat\varrho\in L(\cH)$ is bounded as
\begin{align}
  \|e^{\cL t}\hat\varrho\|_1\le\|e^{\cL t}\|_{\rm op} \|\hat\varrho\|_1\le \|\varrho\|_1,  \forall \varrho \in L(\cH)
\label{bound1}
\end{align}
where we introduced the operator norm
\begin{align}
  \|\cO\|_{\rm op}=\sup\left\{\|\cO \hat x\|_1:\, \hat x\in L(\cH),\, \| \hat x\|_1=1\right\}.
  \label{}
\end{align}

Our goal now is to find a smaller bound than the bound given in Eq. \eqref{bound1} for the 
elements $\hat\varrho(t)=e^{\cL t}\hat\varrho$ but projected to the subspaces with fixed value $|l|$.
For this purpose  we consider the projectors
\begin{align}
\cQ_l=\cP_l+\cP_{-l},\quad l\ne 0,
  \label{}
\end{align}
which act as the identity operator for system $B$. By taking fixed values of $|l|$ one ensures that the  
self-adjoint property is preserved, i.e. $(\cQ_{l}\hat\rho)^\dagger=\cQ_{l}\hat\rho$
whenever $\hat\rho^\dagger=\hat\rho$. This does not hold when projecting with $\cP_{l\neq 0}$.
Therefore $\cQ_l\hat\varrho\in L(\cH)$ and the 
operator norm of each projector $\|Q_l\|_{\rm op}=1$. The projector $\cQ_l$
projects all elements $\hat x \in L(\cH)$ onto a closed linear subspace of $L(\cH)$ which is also
a Banach space with norm $\|.\|_1$:
\begin{align}
  L^l(\cH):=\left\{\cQ_l  \hat x:\, \hat x\in L(\cH) \right\}.
  \label{BanachL1l}
\end{align}

In the next step we take advantage of the fact that the generators $\cL$ and  $\tilde\cL^A$ do not mix spaces of 
different value of $l$, i.e. $\cQ_l\cL=\cL\cQ_l$. This allows
us to define the operators restricted to $L^l(\cH)$
\begin{align}
  \cL_l:=\cQ_l\cL=\cL\cQ_l,\quad
  \tilde\cL^A_l:=\cQ_l\tilde\cL^A=\tilde\cL^A\cQ_l.
  \label{}
\end{align}
Now we can evaluate the norm of the projected elements as
\begin{align}
  \|\cQ_l\hat\varrho(t)\|_1
 \le 
 \|e^{\cL_l t}\|^l_{\rm op}\,\|\cQ_l\hat\varrho(0)\|_1.
  \label{}
\end{align}
where we also used that $\cQ_l^2=\cQ_l$.
The operator norm $\|.\|^l_{\rm op}$ is induced by the Banach space 
$L^l(\cH)$. 
In order to find a better bound we take advantage of our knowledge of the generator
$\tilde \cL^A$ and separate the dynamics with the use of
the Trotter product formula (see Corollary $5.8$ in Chap. III of Ref. \cite{Engel})
\begin{align}
 e^{\cL_l t}=
  \lim_{N\to\infty}
  \left(e^{\tilde\cL_l^A t/N}
  e^{(\cL_l-\tilde\cL_l^A)t/N}\right)^N.
  \label{Trotter}
\end{align}
 
As discussed in Appendix \ref{Appendix0}  the Liouvillian $\tilde\cL^A$ has only a point spectrum . Therefore
in the Banach space $L^l(\cH)$ the dynamical semigroup of system $A$ has 
the operator norm
$\|\exp(\tilde\cL^A_l t)\|^l_{\rm op} \le e^{\eta_l t}$ (see Definition $2.1$ and Corollary $3.12$ in Chap. IV of Ref. \cite{Engel})
where $\eta_l\in\mathbb{R}_-$ is the largest eigenvalue of $\tilde\cL^A_l$ in the 
Banach space $L^l(\cH)$ ($l\ne 0$). 
This follows from the fact that 
there is only one zero eigenvalue of $\tilde \cL^A$ with corresponding eigenvector 
$\hat\rho_{\rm st}^A$ that has no projection on the subspaces with $l\neq 0$, i.e. 
$\cQ_l\hat\rho_{\rm st}^A=0$
and $\cP_0\hat\rho_{\rm st}^A=\hat\rho_{\rm st}^A$. The  rest of
the  eigenvalues of $\tilde\cL^A$ are negative. 
This allows us to find a bound independent of $N$ for the evolution
operator that we expanded using the Trotter formula in Eq. \eqref{Trotter}. Using the submultiplicative property of the operator norm $\|.\|^l_{\rm op}$ and the 
fact that $(\cL_l-\tilde\cL_l^A)$ also generates  a strongly continuous one-parameter contracting semigroup, we have
\begin{eqnarray}
   \|e^{\tilde\cL_l^A t/N}e^{(\cL_l-\tilde\cL_l^A)t/N}\|^l_{\rm op}&\le&  \|e^{\tilde\cL_l^A t/N}\|^l_{\rm op} \|e^{(\cL_l-\tilde\cL_l^A)t/N}\|^l_{\rm op} \nonumber \\
    &\le& \|e^{\tilde\cL_l^A t/N}\|^l_{\rm op} \le e^{\eta_l t/N}
\end{eqnarray}
which yields
\begin{align}
   \| \lim_{N\to\infty}\left(e^{\tilde\cL_l^A t/N}e^{(\cL_l-\tilde\cL_l^A)t/N}\right)^N\|^l_{\rm op}\le e^{\eta_l t}.
\label{}
\end{align}
This upper bound is independent of $N$ which implies the following bound for the norm 
\begin{align}
  ||\cQ_l\hat\varrho(t)||_1\le e^{\eta_l t}||\varrho(0)||_1\quad
  \Rightarrow\quad \lim_{t\to \infty}\cQ_l\hat\varrho(t)=0.
  \label{}
\end{align}
Thus the steady state has projection $0$ on the subspaces $l\ne 0$. This fact
does not depend on the interaction Hamiltonian $\hat V^A \hat V^B$. This is a stronger result than the original requirement for our 
statement, because 
\begin{equation}
 \mathrm{Tr}_B \{\cQ_l\hat\varrho(t\to \infty)\}=0 \Rightarrow \mathrm{Tr}_B \{\hat\varrho(t\to \infty)\}=\hat \rho^A_{\rm st},
\end{equation}
but we found that $\cQ_l\hat\varrho(t\to \infty)=0$. 

The above approach can also be used for system $A$ alone in the case when we know the spectrum of the Liouville operator, but
due to the internal interactions it is hard to solve the dynamics or to find the steady state. As an example someone can think
about the Jaynes-Cummings model where the two-level system and the harmonic oscillator have an independent damping mechanism.
Despite the fact that finding the analytical form of the steady state is a challenging task, in an appropriate representation we have to deal only with the 
block diagonal part of the density matrix.

We end this section with a comment on the choice of separate damping mechanisms. This is a valid description for
weakly interacting systems, where the environments are insensitive to the dressed eigenstates of coupled 
systems $A$ and $B$. For strong coupling, 
the physical description is not valid anymore and a more detailed lossy model
has to be derived. For instance, in Ref. \cite{Hu} this was investigated for the optomechanical model 
and the master equation was derived using the dressed state of the Hamiltonian in Eq. \eqref{optoint}.  
The result is a master equation that differs from Eq. 
\eqref{mainEq} and \eqref{optoL} by an additional dissipative term proportional to $\cD[\hat a^\dagger \hat a]$
and where the mechanical dissipators are replaced by
$\cD[\hat b]\to \cD[\hat b-g/\nu \hat a^\dagger \hat{a}]$ and $\cD[\hat b^\dagger] \to 
\cD[\hat b^\dagger-g/\nu \hat a^\dagger \hat{a}]$. With these corrections, the condition in 
Eq. \eqref{LA} is still fulfilled and therefore
the steady of system $A$, i.e. the optical mode, remains invariant as well. 
In general, our method applies to any two dissipative systems $A$ and $B$ with a separable Hamiltonian interaction
$\hat V^A\hat V^B$ and 
where: the assumptions \eqref{HA} and \eqref{LA} are satisfied, and one can identify a generator
$\tilde \cL^A$ acting solely on system $A$ 
whose zeroth eigenvalue belongs only to the eigenspace $l=0$ of $\cC^A$ (see Eq. \eqref{eigenC}). 
There are techniques to infer about the spectrum of a finite dimensional operator.
This implies that our results can be applied to finite dimensional systems including  
corrections due to the strong coupling limit.
However, in the case of infinite dimensional Hilbert spaces with non-zero temperature baths, 
the choice is restricted because only the spectral decomposition of the optical master equation 
(damped harmonic oscillator) is known.

\section{Examples}
\label{ex}
\subsection{A case study of two interacting spins}

The situation can be easily understood for the simplest case of two interacting spin 
systems ($A$ and $B$) with independent Markovian dynamics. 
We consider an interaction Hamiltonian which satisfies the symmetries of the optomechanical system.
Due to the simple structure of the Hilbert spaces the steady state of the whole system has an analytical
formula. The Hilbert spaces 
are $\mathcal{H}_A=\mathcal{H}_B=\mathbb{C}^2$ with orthonormal basis
$\{\ket{0}_A,\ket{1}_A\}$ for system $A$ and $\{\ket{0}_B,\ket{1}_B\}$ for system $B$. 
Each system alone is governed by the master equation
$d\hat \rho^i/dt=\cL^i\hat\rho^i$ 
with $i\in\{A,B\}$
expressed in terms of the the Liouvillians
\begin{eqnarray}
\label{twolevel}
\mathcal{L}^i\hat{\rho}^i=i\omega[\hat{\rho}^i,\hat{\sigma}_z^i]+\gamma^i(1-s^i)
\mathcal{D}[\hat{\sigma}^i_-]\hat{\rho}^i 
+\gamma^i s^i \mathcal{D}[\hat{\sigma}^i_+]\hat{\rho}^i 
\end{eqnarray}
where $\omega^i$ is the frequency difference between the energy levels $\ket{0}_i$ and $\ket{1}_i$. $\gamma^i$ has 
the dimension of frequency and $0\leq s^i\leq 1$. The spin operators are given by 
\begin{eqnarray}
\hat{\sigma}^i_z=\ket{1}_i\bra{1}_i-\ket{0}_i\bra{0}_i, \quad 
\hat{\sigma}^i_-=(\hat{\sigma}^i_+)^\dagger=
\ket{0}_i\bra{1}_i.
\end{eqnarray}
The steady state of each separate system has the same form
\begin{equation}
 \hat{\rho}^i_{\rm st}=(1-s^i)\ket{0}_i\bra{0}_i+s^i\ket{1}_i\bra{1}_i.
\label{freesteady1}
\end{equation}
By considering the interaction Hamiltonian 
\begin{equation}
\hat{H}_I= \Omega \hat{\sigma}^A_z\left(\hat{\sigma}^B_++\hat{\sigma}^B_-\right), 
\end{equation}
one can realize that the interaction Hamiltonian commutes with the
free Hamiltonian of system $A$, i.e.
\begin{eqnarray}
 &&[\hat{H}_I,\hat{\sigma}^A_z]=0.
\end{eqnarray}
The steady state $\hat\varrho_{\rm st}$ of the whole system can be analytically evaluated due
to the small dimension of the Hilbert space. We do not show the explicit expressions here but simply emphasize that 
it has  the form
\begin{equation}
\hat\varrho_{\rm st}=\left(
  \begin{array}{cccc}
   \rho_{0,0,0,0}&\rho_{0,0,0,1}&0&0\\
    \rho_{0,1,0,0}&\rho_{0,1,0,1}&0&0\\
    0&0&\rho_{1,0,1,0}&\rho_{1,0,1,1}\\
    0&0&\rho_{1,1,1,0}&\rho_{1,1,1,1}
  \end{array}
  \right), 
\end{equation}
where $\rho_{i,j,k,l}=\bra{i}_A \bra{j}_B\hat\varrho_{\rm st}\ket{k}_A \ket{l}_B$ and $i,j,k,l \in \{0,1\}$. 
The reduced steady state of system $A$ is given by
\begin{eqnarray}
 \mathrm{Tr}_B\{\hat\varrho_{\rm st}\}=\left(\begin{array}{cc}
   1-s^A&0\\
    0&s^A
  \end{array}\right)
=\hat{\rho}^A_{\rm st},
\end{eqnarray}
whereas for system $B$ it reads
\begin{equation}
 \mathrm{Tr}_A\{\hat\varrho_{\rm st}\}=\left(\begin{array}{cc}
   B_1&B_2\\
    B^*_2&1-B_1
  \end{array}\right)\neq\hat{\rho}^B_{\rm st}.
\label{formofB} 
\end{equation}
The steady state of system $B$ is given by the real parameter $B_1$ and the complex parameter $B_2$ both of them being
functions of $\omega$, $\Omega$, $\gamma^{i}$ and $s^{i}$ ($i \in \{A,B\}$). In the case of strong interaction between
systems $A$ and $B$ the following relations hold
\begin{equation}
\lim_{\Omega \to \infty}B_1=1/2,\,\,\lim_{\Omega \to \infty}B_2=0,
\end{equation}
resulting in a maximally mixed state for $\mathrm{Tr}_A\{\hat{\varrho}_{\rm st}\}$. 
\begin{figure}[h]
\begin{center}
  \includegraphics[width=8.cm]{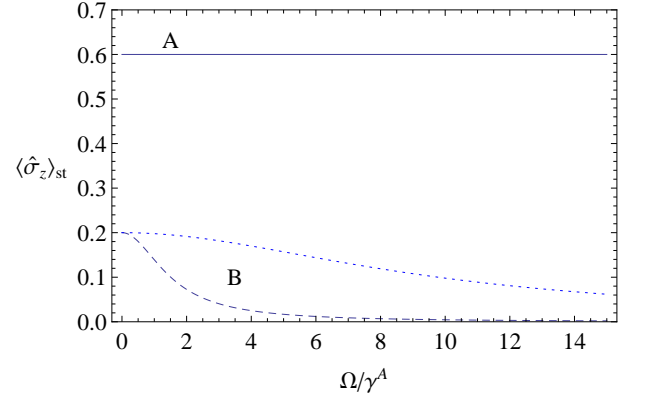}
  \caption{\label{fig1} The population inversion of system $A$ and $B$ in the steady state as a function of $\Omega/\gamma^A$. 
We set $\gamma^A/\gamma^B=1$, $s^B=0.6$ and
$s^A=0.8$. The solid line 
shows that in system $A$ the average excitation difference between the energy levels $\ket{1}_A$ and $\ket{0}_A$ remains independent of interaction strength $\Omega$, while 
the dashed and the dotted line related to system $B$ shows a decrease.
The parameters are $\omega/\gamma^A=1$ (dotted curve) and $\omega/\gamma^A=10$ (dashed curve).}
\end{center}
\end{figure} 

In Fig. \ref{fig1} we found that the population inversion $\langle \hat{\sigma}^B_z \rangle_{\rm st}= \mathrm{Tr}\{ \hat{\sigma}^B_z\hat\varrho_{\rm st}\}$ of system $B$ has a monotone 
decreasing behaviour as a function of the interaction strength $\Omega$, and furthermore its 
dependence on the frequency difference between the energy levels $\omega$ governs the rate of the decrease. 
The decrease is slowed down by the increase of the frequency difference $\omega$. System $A$ returns to the same steady state 
independently of the interaction strength $\Omega$. The steady state of system $B$ depends on the interaction strength which
can outplay the relaxation mechanism of system $B$. This preliminary result already hints that this model can only explain physical situations whereas $\Omega \ll \omega$.  

\subsection{A spin system interacting with a harmonic oscillator}
\label{ex2}

In this section we consider an interaction Hamiltonian between a finite and an infinite dimensional system. 
The Hilbert spaces are $\mathcal{H}_A=\mathbb{C}^2$ and $\mathcal{H}_B=\Gamma_s(\mathbb{C})$ 
(symmetric Fock space) with orthonormal basis
$\{\ket{0}_A,\ket{1}_A\}$ for system $A$ and $\{\ket{n}_B\}_{n \in \mathbb{N}}$ for system $B$. System $A$ is governed by the master equation 
\begin{eqnarray}
\frac{d\hat{\rho}^A}{dt}&=&\mathcal{L}^A\hat{\rho}^A \\
&=&i\omega_A[\hat{\rho}^A,\hat{\sigma}_z]+\gamma^A (1-s)\mathcal{D}[\hat{\sigma}_-]\hat{\rho}^A+\gamma^A s \mathcal{D}[\hat{\sigma}_+]\hat{\rho}^A,   \nonumber
\end{eqnarray}
where we use the spin operators $\hat{\sigma}_z=\ket{1}_A\bra{1}_A-\ket{0}_A\bra{0}_A$ and $\hat{\sigma}_-= \hat{\sigma}^\dagger_+=\ket{0}_A\bra{1}_A$.
The parameter $\omega_A$ stands for the frequency difference between the levels $\ket{1}_A$, $\ket{0}_A$, $\gamma^A$ has the dimension of
frequency and $0\leq s \leq1$. System $B$ is governed by the master equation 
\begin{eqnarray}
\frac{d\hat{\rho}^B}{dt}&=&\mathcal{L}^B\hat{\rho}^B \\
&=&i\omega_B[\hat{\rho}^B,\hat{b}^\dagger \hat{b}]+\gamma^B (1+\bar n)\mathcal{D}[\hat{b}]\hat{\rho}^B+\gamma^B \bar n \mathcal{D}[\hat{b}^\dagger]\hat{\rho}^B, \nonumber 
\end{eqnarray}
where $\hat{b}$ ($\hat{b}^\dagger$) is the annihilation (creation) operator of the harmonic oscillator with frequency $\omega_B$. 
$\bar n \geq 0$ is proportional to the temperature of the thermal reservoir to which the harmonic oscillator is coupled and $\gamma^A$ has the dimension of
frequency.

The steady states of the two systems are (see Appendix \ref{Appendix})
\begin{eqnarray}
 \hat{\rho}^A_{\rm st}&=&(1-s)\ket{0}_A\bra{0}_A+s\ket{1}_A\bra{1}_A, 
\nonumber \\
 \hat{\rho}^B_{\rm st}&=&\frac{1}{\bar n+1} \sum^\infty_{n=0} \left(\frac{\bar n}{\bar n +1}\right)^n \ket{n}_B \bra{n}_B.
\end{eqnarray}
We assume that the two systems interact through the Hamiltonian
\begin{equation}
 \hat{H}_I= \Omega \hat{\sigma}_z\left(\hat{b}^\dagger+\hat{b}\right),
\label{HI4}
\end{equation}
and the evolution of the whole system is given by the master equation
\begin{equation}
 \frac{d \hat \varrho}{dt}=-i [\hat{H}_I, \hat \varrho]+ \mathcal{L}^A \hat \varrho + \mathcal{L}^B \hat \varrho.
 \label{twolevel2}
\end{equation}
A similar type of optical master equation has been suggested to describe the properties 
of defect centers in crystalline structures \cite{Betzholz}.
The eigenvectors of $\hat V^A=\hat{\sigma}_z$ are $\ket{0}_A$ and $\ket{1}_A$ and therefore we can construct the eigenvectors of the superoperator $\cC^A \hat \rho=
[\hat{\sigma}_z,\hat \rho]$:
\begin{eqnarray}
 &&\hat \sigma_{l=1,\nu=0}=\ket{0}_A \bra{1}_A,\,\, \hat \sigma_{l=-1,\nu=0}=\ket{1}_A \bra{0}_A, \nonumber \\
 &&\hat \sigma_{l=0,\nu=0}=\ket{0}_A \bra{0}_A,\,\, \hat \sigma_{l=0,\nu=1}=\ket{1}_A \bra{1}_A.
\end{eqnarray}
In the case when $l=\pm 1$ the Liouville operator $\tilde \cL^A \hat \rho =\cL^A \hat \rho + i \omega_A [\hat \sigma_z, \hat \rho]$ has only one eigenvalue
$\eta_{l= \pm 1}= - \gamma^A /2$.

It follows from the results of Sec. \ref{ststa} that the steady state of the two-level system remains independent of the interaction strength $\Omega$. However,
for this particular case we are going to show the independence by a direct calculation. Therefore, in order to investigate the interaction of 
the two systems and also their independent damping mechanism we now consider the density matrix
in the following representation
\begin{equation}
\hat{\varrho}=\int \left(\begin{array}{cc}
   P_{0,0}(\alpha,\alpha^*)& P_{0,1}(\alpha,\alpha^*)\\
   P_{1,0}(\alpha,\alpha^*)&P_{1,1}(\alpha,\alpha^*)
  \end{array}\right) \ket{\alpha}_B\bra{\alpha}_B \, d^2 \alpha,
\label{repres2}
\end{equation}
where $d^2 \alpha=d\mathrm{Re}\{\alpha\}\,d\mathrm{Im}\{\alpha\}$ and $P_{i,j}(\alpha,\alpha^*)$($i,j \in \{0,1\}$) are quasi-distributions
of Glauber-Sudarshan type \cite{Glauber,Sudarshan}. We are going to focus only on the off-diagonal element $P_{0,1}(\alpha,\alpha^*)$ and using
the following identities
\begin{equation}
 \hat{b}^\dagger \left(\ee^{\frac{|\alpha|^2}{2}}\ket{\alpha}_B\right) = 
 \frac{\partial}{\partial \alpha} \left(\ee^{\frac{|\alpha|^2}{2}}\ket{\alpha}_B\right),\,\, 
\hat{b} \ket{\alpha}_B= \alpha \ket{\alpha}_B,
\label{GSidentities}
\end{equation}
the equation of motion \eqref{twolevel2} for the representation \eqref{repres2} yields
\begin{eqnarray}
\frac{dP_{0,1}}{dt}&=&\left(2i \omega_A- \frac{\gamma^A}{2} \right) P_{0,1}+D[P_{0,1}] \nonumber \\
         &-&i \Omega \left(\frac{\partial}{\partial \alpha^*}+\frac{\partial}{\partial \alpha}-2\alpha^*-2\alpha \right)P_{0,1},
\label{1}
\end{eqnarray}
where
\begin{eqnarray}
&& D[P_{0,1}]=-i\omega_B \left(\frac{\partial}{\partial \alpha^*}\alpha^*-\frac{\partial}{\partial \alpha}\alpha\right)P_{0,1}   \\
&&+\frac{\gamma^B}{2}\Big[\left(\frac{\partial}{\partial \alpha^*}\alpha^*+\frac{\partial}{\partial \alpha}\alpha\right)+
2 \bar n \frac{\partial^2}{\partial \alpha \partial \alpha^*}\Big]P_{0,1}. \nonumber
\end{eqnarray}
In order to obtain the above differential equations an integration by parts was taken in the 
variables $\alpha$ and $\alpha^*$. The structure of Eq. \eqref{1} suggests a Gaussian ansatz of the form
\begin{equation}
\label{2}
P_{0,1}=\ee^{-a(t)+b(t)\alpha+c(t)\alpha^*-d(t)\mid \alpha \mid^2}.
\end{equation}
Substituting Eq. \eqref{2} into the Eq. \eqref{1}, and a comparison of 
the terms proportional to $1$, $\alpha$, $\alpha^*$ and $|\alpha|^2$ leads to the following set of differential equations:
\begin{eqnarray}
\dot{a}&=&-2i\omega_A-i\Omega\left(c+b\right)-\gamma^B \bar n\left(bc-d\right) 
             -\gamma^B+\gamma^A/2, \nonumber \\ 
\dot{b}&=&i \omega_B b-i\Omega\left(d+2\right)+\gamma^B/2 \,b
             - \gamma^B \bar n \, bd, \nonumber \\
\dot{c}&=&-i \omega_B c-i\Omega\left(d+2\right)+\gamma^B/2 \, c
             - \gamma^B \bar n \, cd, \nonumber \\
 \dot{d}&=&\gamma^B d-\gamma^B \bar n \,d^2. 
\end{eqnarray}
The solution of these equations is very complicated and is subject to the initial conditions. However, we are interested in the limit $t \to \infty$  and 
$b(t)$, $c(t)$ and $d(t)$ tend to a finite limit, but the normalization of Eq. \eqref{2} is characterized by $a(t) \approx 
\gamma^At/2$. Therefore in the long time limit $ P_{0,1}(t \to \infty)=0$ and the decay to this limit is upper bounded by $\ee^{-\gamma^At/2}$. The same arguments apply
to $P_{1,0}(t)$. It is immediate that the steady state of the spin system remains invariant
\begin{equation}
 \mathrm{Tr}_B\{\hat \varrho(t \to \infty)\}=(1-s)\ket{0}_A\bra{0}_A+s\ket{1}_A\bra{1}_A.
\end{equation}

In order to gain some insight
into the evolution of the diagonal elements, we study the average excitation of the harmonic oscillator 
$\langle \hat{b}^\dagger \hat{b} \rangle=\mathrm{Tr}\{\hat{b}^\dagger \hat{b} \hat{\varrho}(t)\}$ and the population inversion of the 
spin system $\langle \hat{\sigma}_z \rangle=\mathrm{Tr}\{\hat{\sigma}_z \hat{\varrho}(t)\}$.
Making use of Eq. \eqref{twolevel2} we can derive the following equations of motion for $\langle \hat{b}^\dagger \hat{b} \rangle$ and
for $\langle \hat{\sigma}_z \rangle$
\begin{eqnarray}
 \frac{d}{dt}\langle \hat{b}^\dagger \hat{b} \rangle&=&
-2\Omega \,\mathrm{Im}[\langle \hat{\sigma}_z\hat{b}  \rangle]-\gamma^B \langle \hat{b}^\dagger \hat{b} \rangle +\gamma^B \bar n \nonumber \\
 \frac{d}{dt} \langle \hat{\sigma}_z\hat{b}  \rangle&=&\left(-i \omega_B-\frac{\gamma^B}{2}-\gamma^A \right)
\langle \hat{\sigma}_z\hat{b}  \rangle-i\Omega \nonumber \\
&&+\gamma^A\left(2s-1\right)\langle \hat{b} \rangle, \nonumber \\
\frac{d}{dt} \langle \hat{b}  \rangle&=&\left(-i\omega_B-\frac{\gamma^B}{2}\right)\langle \hat{b}  \rangle-i\Omega 
\langle \hat{\sigma}_z\rangle, \nonumber \\
 \frac{d}{dt} \langle \hat{\sigma}_z  \rangle&=&-\gamma^A \langle \hat{\sigma}_z  \rangle+ \gamma^A (2s-1),
\label{exceq1}
\end{eqnarray}
where $\mathrm{Im}[z]$ denotes the imaginary part of the complex number $z$.

The obtained equations form a closed set of differential equations and they can be solved for a set of given initial conditions. However,
we are interested in the behaviour of these quantities in the steady state $\hat{\varrho}_{\rm st}$. Setting the left hand side of 
Eq. \eqref{exceq1} to zero we find the following solutions to the population inversion
\begin{equation}
\langle \hat{\sigma}_z  \rangle_{\rm st}=\mathrm{Tr}\{\hat{\sigma}_z \hat{\varrho}_{\rm st}\}=2s-1, 
\end{equation}
and to the average excitation of the harmonic oscillator
\begin{eqnarray}
&&\langle \hat{b}^\dagger \hat{b}  \rangle_{\rm st}
=
\bar n + 
(2s-1)^2 \frac{4\Omega^2}{\left( \gamma^B \right)^2+4 \omega^2_B} 
\nonumber \\
&&+  4(1-s)s\frac{2 \gamma^A+\gamma^B}{ \gamma^B}\frac{4\Omega^2}{\left(2 \gamma^A+\gamma^B\right)^2+4 \omega^2_B}.
\label{excosc}
\end{eqnarray}
 The average excitation of the harmonic oscillator $\langle \hat{b}^\dagger \hat{b}  \rangle_{\rm st}$ shows an increasing dependency on the interaction strength $\Omega$.
The most peculiar scenario occurs when $s=\bar n=0$, in which case the non-interacting systems relax to their ground state
$\hat{\rho}^i_{\rm st}=\ket{0}_i \bra{0}_i$ ($i \in \{A,B\}$). While this excitation extracting mechanism dominates both systems, due to the interaction Hamiltonian in Eq. \eqref{HI4}
the average excitation of the harmonic oscillator for $\Omega >0$ takes the value
\begin{equation}
\langle \hat{b}^\dagger \hat{b}  \rangle_{\rm st}=\frac{4 \Omega^2}{\left(\gamma^B\right)^2+4 \omega^2_B}. 
\end{equation}
\begin{figure}[h]
\begin{center}
  \includegraphics[width=8.cm]{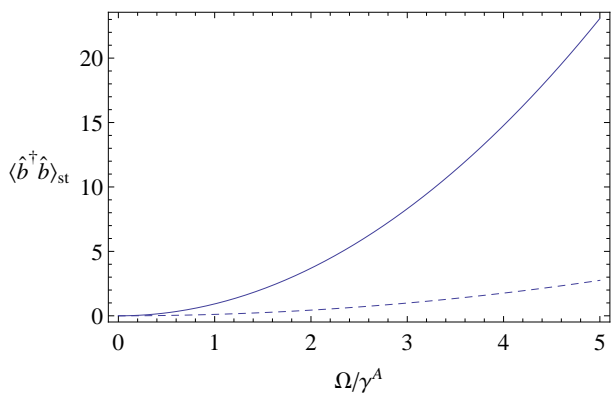}
  \caption{\label{fig2} The average excitation of the harmonic oscillator in the steady state as a function of $\Omega/\gamma^A$. 
We set $\gamma^B/\gamma^A=1$, $\bar n=0$ and $s=1/2$. The solid line with $\omega^B/\gamma^A=1$ and the dashed line with $\omega^B/\gamma^A=5$ 
shows an increase in the average excitation, but this increase is mediated by the frequency of the harmonic oscillator $\omega^B$. }
\end{center}
\end{figure} 

Fig. \ref{fig2} shows that the average excitation of the harmonic oscillator starts at the value $\bar n$ and increases as a function of 
$\Omega$. This increase depends on the relaxation parameters of the spin system and of the harmonic oscillator. The found behaviour suggests that the model
can only describe physical scenarios whenever $\omega^B \gg \Omega \sqrt{\gamma^A/ \gamma^B}$.

\begin{figure}[h]
\begin{center}
  \includegraphics[width=8.cm]{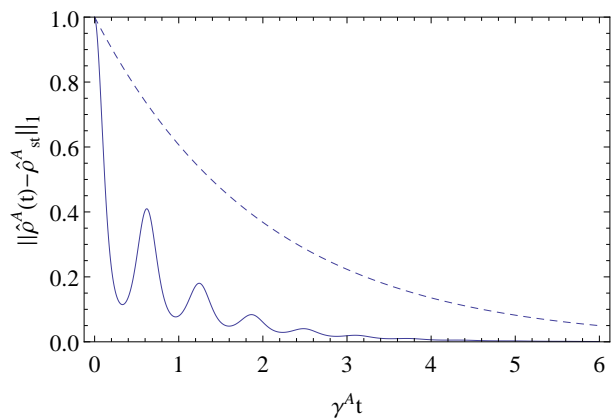}
  \caption{\label{fig3} The norm distance between the evolved state $\hat \rho^A(t)=\mathrm{Tr}_B\{\hat \varrho(t)\}$ and the steady state of system $A$
  as a function of $\gamma^A t$. We set $\gamma^B/\gamma^A=1$, $\bar n=0$, $s=1/2$ and $\omega^A/\gamma^A=\omega^B/\gamma^A=10$. The solid line with $\Omega/\gamma^A=5$ 
  and the dashed line with $\Omega/\gamma^A=0$ shows the difference between coupled and uncoupled system.}
\end{center}
\end{figure} 

We also investigated numerically the decaying mechanism of the whole system. We considered the initial condition to be $\frac{\ket{0}_A+\ket{1}_A}{\sqrt{2}} \ket{0}_B$.
Thereby choosing $\bar n=0$ and $s=1/2$ the only elements affected by the dynamics are the off-diagonal elements $~_A\bra{0} \hat \varrho(t) \ket{1}_A$ and 
$~_A\bra{1} \hat \varrho(t) \ket{0}_A$. We simulated the harmonic oscillator with ten basis vectors which are enough because the initial condition is the steady state
for system $B$ and the chosen coupling strength $\Omega$ is not big enough to drive the harmonic oscillator towards higher excitations than $10$ and the contributions
from these states are negligible. Fig. \ref{fig3} shows that the decay $\ee^{-\gamma^At/2}$ of the uncoupled system $A$, i.e., $\Omega=0$, is the upper bound for 
the off-diagonal elements which is in agreement with our general findings in Sec. \ref{ststa}.

\subsection{The optomechanical system}
\label{ex3}

In this section we study more in detail
the model which motivated our work, see Sec. \ref{ex1}. The system is composed of 
two harmonic oscillators, i.e., two infinite systems, and a direct approach
to find out the steady state of the joint system requires the solutions of an infinite number of recurrence relations. Therefore our result in Sec. \ref{ststa} is useful to
identify some properties of the steady state. The Hilbert spaces are 
$\mathcal{H}_A=\Gamma_s(\mathbb{C})$ and $\mathcal{H}_B=\Gamma_s(\mathbb{C})$ 
 with orthonormal basis $\{\ket{n}_A\}_{n \in \mathbb{N}}$ for system $A$ 
 (optical mode) and $\{\ket{n}_B\}_{n \in \mathbb{N}}$ for system $B$
 (mechanical mode). The uncoupled systems are governed by the master equation
\begin{equation}
\label{optomech}
\frac{d\hat{\rho}^A}{dt}=\cL^A\hat{\rho}^A,\quad 
\frac{d\hat{\rho}^B}{dt}=\cL^B\hat{\rho}^B, 
\end{equation}
with the Liouville operators defined in Eq. \eqref{optoL}.
We recall that $\hat{a}$ ($\hat{a}^\dagger$) and $\hat{b}$ ($\hat{b}^\dagger$)are the annihilation (creation) operators of the quantum harmonic 
oscillators $A$ and $B$ with frequency $\omega$ and frequency $\nu$ respectively. 
The coefficients with dimensions of frequency
characterizing the dissipative dynamics are  $\kappa$  and $\gamma$
for the optical and mechanical mode respectively.
The temperature of the photonic reservoir is proportional to $\bar n$ and
to $\bar m$ in the phononic case.
The steady states of the separate systems are (see Appendix \ref{Appendix})
\begin{eqnarray}
 \hat{\rho}^A_{\rm st}&=&\frac{1}{\bar n + 1} \sum^\infty_{n=0} \left(\frac{\bar n}{\bar n + 1}\right)^n \ket{n}_A \bra{n}_A, 
\nonumber \\
 \hat{\rho}^B_{\rm st}&=&
 \frac{1}{\bar m + 1} \sum^\infty_{n=0} \left(\frac{\bar m}{\bar m + 1}\right)^n \ket{n}_B \bra{n}_B.
 \label{optosteady}
\end{eqnarray}
The two systems interact with each other through the optomechanical interaction
$ \hat{H}_I= g \hat{a}^\dagger \hat{a} \left(\hat{b}^\dagger+\hat{b}\right)$
and the evolution of the whole system is given by the master equation \eqref{mainEq}.

Connecting with the formalism in Sec. \ref{ststa} we note that in
this case $\hat V^A=\hat{a}^\dagger \hat{a}$ and for all $n \in \mathbb{N}$ $\ket{n}_A$ is an eigenvector with eigenvalue $n$. The eigenvectors of the superoperator $\cC^A \hat \rho=
[\hat{a}^\dagger \hat{a},\hat \rho]$ are
\begin{equation}
 \hat \sigma_{l,n}=\begin{cases} \ket{n+l}_A \bra{n}_A, \, l\geq 0 \\
                    \ket{n}_A \bra{n+l}_A, \, l< 0 \\
                   \end{cases}  n \in \mathbb{N}.
\end{equation}
In the subspaces defined by the eigenvectors $l \neq 0$ the smallest eigenvalue is $\eta_l=-\kappa |l|/2$ (see Appnedix \ref{Appendix0}). Applying the results of
Sec. \ref{ststa} we get
\begin{equation}
\mathrm{Tr}_B\{\hat{\varrho}_{\rm st}\}=\hat{\rho}^{A}_{\rm st}.
\end{equation}
In this particular example it is hard to work out an alternative proof for the above relation as we did in section \ref{ex2}. In general 
$\mathrm{Tr}_A\{\hat{\varrho}_{\rm st}\} \neq \hat{\rho}^{B}_{\rm st}$ and using the same approach as in the previous section we
are going to analyze the average excitations  
$\langle \hat{a}^\dagger \hat{a} \rangle=\mathrm{Tr}\{\hat{a}^\dagger \hat{a} \hat{\varrho}(t)\}$ and 
$\langle \hat{b}^\dagger \hat{b} \rangle=\mathrm{Tr}\{\hat{b}^\dagger \hat{b} \hat{\varrho}(t)\}$.
The time evolution of the density matrix $\hat{\varrho}(t)$ is governed by Eq. \eqref{optoint} and the following equations of motion can be derived 
for $\langle \hat{a}^\dagger \hat{a} \rangle$ and $\langle \hat{b}^\dagger \hat{b} \rangle$
\begin{eqnarray}
\label{twoharm}
\frac{d}{dt}\langle \hat{b}^\dagger \hat{b} \rangle&=&
-2 g \, \mathrm{Im}[\langle \hat{a}^\dagger \hat{a} \hat{b} \rangle]-\gamma \langle \hat{b}^\dagger \hat{b} \rangle + \gamma \bar m, \nonumber \\
\frac{d}{dt}\langle \hat{a}^\dagger \hat{a} \hat{b} \rangle&=&
\kappa \bar n \langle \hat{b} \rangle 
-\frac{\gamma+2(\kappa+i\nu)}{2} \langle \hat{a}^\dagger \hat{a} \hat{b} \rangle
-ig \langle \left(\hat{a}^\dagger \hat{a} \right)^2\rangle
\nonumber \\
\frac{d}{dt}\langle \hat{b} \rangle&=&-i \nu \langle \hat{b} \rangle - i g \langle \hat{a}^\dagger \hat{a}\rangle
-\frac{\gamma}{2} \langle \hat{b} \rangle, \nonumber \\
\frac{d}{dt}\langle \left(\hat{a}^\dagger \hat{a}\right)^2 \rangle&=&
\kappa \bar n
-2\kappa
\langle \left(\hat{a}^\dagger \hat{a}\right)^2 \rangle+\kappa(4 \bar n +1)\langle\hat{a}^\dagger \hat{a} \rangle, \nonumber \\
\frac{d}{dt}\langle \hat{a}^\dagger \hat{a} \rangle&=&-\kappa
\langle \hat{a}^\dagger \hat{a}\rangle + \kappa \bar n.
\end{eqnarray}
The steady state solutions of Eq. \eqref{twoharm}  are
\begin{eqnarray}
\label{st2harm}
\langle \hat{a}^\dagger \hat{a} \rangle_{\rm st}&=&\bar n, \\
\langle \hat{b}^\dagger \hat{b} \rangle_{\rm st}&=&\bar m+
 \frac{4 \bar n^2 g ^2}{\gamma^2+4\nu^2}
   +\frac{4 \bar n (\bar n+1)
   (2 \kappa+\gamma) g ^2}{\gamma \left[ ( 2\kappa+\gamma)^2+4 \nu^2\right]}. \nonumber
\end{eqnarray}
The above relation shows that the average excitation of system $B$ is increasing as a function of $g$. The only case when it remains the same is for $\bar n=0$, i.e.,
system $A$ is relaxed to the ground state. So, whenever system $A$ is relaxed to a finite temperature Gibbs state the average excitation number of system $B$ is increased
above the average excitation given by its own damping mechanism. These results resemble the same effect found in Sec. \ref{ex2}. The parameter region where this model
works is constrained by the relation $\nu \gg g  \bar n \sqrt{\kappa/ \gamma}$. 
This is usually fulfilled, because in experimental scenarios $\bar n \approx 0$
\cite{Aspelmeyer}.

\begin{figure}[h]
\begin{center}
  \includegraphics[width=8.cm]{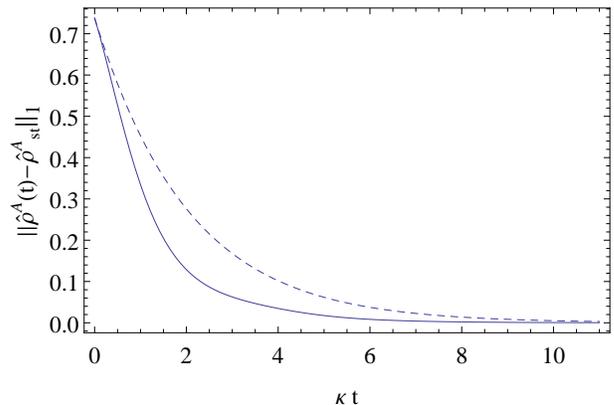}
  \caption{\label{fig4} 
  The norm distance between the evolved 
  state $\hat \rho^A(t)=\mathrm{Tr}_B\{\hat \varrho(t)\}$ and the steady state of system $A$
  as a function of $\kappa t$. The dashed curve corresponds to the 
  free evolution of $A$, i.e. $g=0$. The full line with $g=0.9\kappa$ shows the difference between the coupled and the uncoupled system.
  The initial state for both systems $A$ and $B$ is a coherent state
  $\ket\alpha$, with $\alpha=0.15$. The parameters are: $\omega=10\kappa$, $\nu=1.5\kappa$,
  $\gamma=0.9\kappa$, $\bar n=0.015$ and $\bar m=0.1$.}
\end{center}
\end{figure} 

Fig. \ref{fig4} shows the relaxation of the optical mode
to the steady state. We numerically evaluate 
the trace norm $\|\hat\rho^A(t)-\hat\rho^A_{\rm st}\|_1 $
to measure the distance of the evolved 
state to the steady state in Eq. \ref{optosteady}. 
The dashed line shows the free scenario where the optical mode is solely damped
from an initial coherent state $\ket\alpha$ with $\alpha=0.15$.
The full line shows the reduced density matrix of the optical field after the optomechanical
interaction. Both modes are taken initially in a coherent state with $\alpha=0.15$.
One can note that the trace norm vanishes for both situations, meaning that the
optical system always relaxes to the steady state $\hat\rho^A_{\rm st}$. Even though
the optical  steady state is the same, the relaxation mechanism changes due to the 
interaction with the mechanical mode.

\section{Concluding remarks}
\label{Remarks}

We have discussed the steady state of two interacting quantum systems each of them having an independent non-unitary Markovian evolution. The interaction between these two systems
was considered in a way that it commutes with  one of the system Hamiltonians and that the unitary operator generated by it leaves the Liouvillian of the non-unitary 
dynamics invariant. It is demonstrated that the steady state of the whole system has an 
interesting property, the trace over the other system's degrees of freedom results in a steady state for the commuting system which is also the steady state of 
this system without the interaction. In plain words, the commuting system relaxes always to the same steady state, independently of the interaction strength. 

The method applied here is based on the knowledge of the spectrum of the commuting system's Liouville operator. 
The eigenvalue equations of these generators are
known only in a few cases. Therefore we restricted our statement to the case where 
the system with the commuting properties is composed of spins and harmonic oscillators
with damping mechanisms defined in Eqs. \eqref{Lharm}, and \eqref{Lspin}, though the presented method is general enough to deal with other systems as well. 
In the case when this system is finite (more than two levels) the solutions to the eigenvalue equations can be done 
either with analytical formulas or with numerical simulations. However, we chose the model of a two-level (spin) system with decay processes due to its frequent application in
various physical scenarios. In the case of infinite systems our knowledge is restricted to the harmonic oscillator. 
The decoherence mechanism of a harmonic oscillator can be described 
by an optical master equation whose eigenvalue equations are known \cite{BE}. This is the reason why this model is our second choice. These damping models with their Liouvillian generators
are considered in the decomposition of the commuting system, although we allow internal interactions between the systems. For example: in the case of a spin and a harmonic oscillator
we have the two independent damping mechanisms and a Jaynes-Cummings type interaction between them; for two spins and a harmonic oscillator we have three independent damping mechanisms and 
a Tavis-Cummings type interaction between them; for a spin and two harmonic oscillators we have again three independent damping mechanisms and a dipole coupling in the Lamb-Dicke regime,
an interaction used to describe trapped ions with center-of-mass motion inside an optical cavity. In our analysis we do not have to specify the other system. We comment here that
in the case of a Caldeira-Leggett master equation \cite{Caldeira} the solution to the eigenvalue equation is still unknown.
The proof uses theorems from the theory of one-parameter dynamical semigroups. The reason to apply these results is the Liouville operator of the
harmonic oscillator which is an unbounded operator acting on the Banach spaces of trace-class operators. 
We use the fact that dynamical semigroups are contractive
semigroups on the subspace of trace-class operators whose elements are self-adjoint operators. 
We slice up this subspace into further closed subspaces and derive in each of them
an inequality for the norm of the evolved density matrix with the help of the Trotter product formula. 
These inequalities allow us to identify which elements of the density matrix
tend to zero in the limit of infinite time. 
The projections which splice up the subspace are defined by those vectors which 
are eigenvectors of the interaction's superoperator. 

We considered three examples of this scenario: a spin-spin, a spin-harmonic oscillator, and a two harmonic oscillator system. 
In the first example every detail about the time evolution and the steady state 
can be worked out analytically due to the small dimension of the Hilbert space.
For the other two systems the situation is more complicated and in order to understand the evolution of the system 
we derived closed differential equations for the averaged excitations (harmonic oscillator) 
and the population inversion (spin). 
We focused on the steady state solutions which show that the interaction increases 
the average excitation number only for the system without the commuting properties. 
In the case of the optomechanical model, it turns out that this effect is undetectable in current experimental realizations
due to the weak coupling between the optomechanical resonator and the single mode of the radiation field
and the low value of thermal photons in optical frequencies.

Finally let us make some comments on the limitations and possible extensions 
of our approach. The results of this work are based on a class of coupled systems with
independent Markovian master equations, models which can explain certain physical scenarios like the actual optomechanical experiments. However, it is known
that strong interaction between two coupled  dissipating subsystems leads to serious defects,
whenever independently derived non-unitary evolutions are added to the unitary evolution 
\cite{Walls, Schwendimann, Carmichael, Gardiner, Cresser, Zoubi, Scala, Blais, Agarwal}.
If these defects are present in the solutions, then one must start the theoretical aproach on a microscopic level and derive a new dissipative model. 
The structure of these master equations depends on the interaction between the two subsystems. 
For example, in Ref. \cite{Walls} the two subsystems are harmonic oscillators and their creation and annihilation operators
are mixed up in the generator of the master equation, which leads to an open problem of determining the spectrum of this Liouvillian. 
Our statements are based on the requirement of independent master equations in order to keep the non-commuting system 
arbitrary, however when the 
details of this system are known and they fulfill the assumptions of Sec. \ref{ststa} then the steady state of the commuting system remains invariant,
see for example the strongly coupled dissipative optomechanical model \cite{Hu}.

As a final word we think that the presented analysis may offer interesting perspectives for actual and future models with Markovian master equations.  

\section*{Acknowledgement}

This work is supported by the BMBF project Q.com. We are grateful to 
F. Sokoli, N. Trautmann, H. Frydrych, and G. Alber for stimulating discussions.

\appendix
\section{Eigensystem of the Liouville operator}
\label{Appendix0}
In this appendix we present the eigensystem of a damped two-level system and a damped harmonic oscillator. 
Due to the fact that the Hilbert space of a harmonic oscillator
is infinite dimensional we are going to summarize some basic facts about 
the adjoint of a Banach space operator, a notion needed for the definition of the eigenvectors. 

\subsection{Eigensystem of $\cL_{\rm S}$}
In the main text we considered the following master equation for a spin system
\begin{align}
  \cL_{\rm S} \hat \rho= \gamma (1-\bar m)& 
\left(\hat \sigma_- \hat \rho \hat\sigma_+-\tfrac{1}{2}
\left\{\hat\sigma_+ \hat\sigma_- ,\hat \rho\right\}
\right) 
\nonumber \\
  + \gamma \bar m &
\left(\hat \sigma_+ \hat \rho \hat\sigma_--\tfrac{1}{2}
\left\{\hat\sigma_- \hat\sigma_+ ,\hat \rho\right\}
\right).
\end{align}
The corresponding Liouville operator is non-Hermitian, but it can still 
be diagonalized leading to the dual eigenvalue equation 
\begin{align}
\cL_{\rm S}\hat\rho_i=\lambda_i\hat\rho_i,\quad \cL_{\rm S}^\dagger\check\rho_i=
\lambda^\ast_i\check\rho_i.
\label{}
\end{align}
The elements $\hat\rho_i$ are the four right eigenvectors given by
\begin{align}
\hat{\rho}_{0}=\tfrac{1}{2} \hat{\mathbb I}+\tfrac{2\bar m-1}{2}\hat{\sigma}_z,
\quad \hat\rho_z=\hat{\sigma}_z,\quad 
\hat{\rho}_\pm=\hat{\sigma}_\pm,
\end{align}
with the corresponding eigenvalues
\begin{align}
\lambda_0=0
,\quad,\quad\lambda_z=-\gamma,\quad \lambda_\pm =\-\gamma/2.
\label{}
\end{align}
The adjoint of the Liouville operator is
\begin{align}
  \cL_{\rm S}^\dagger \hat \rho= \gamma (1-\bar m)& 
\left(\hat \sigma_+ \hat \rho \hat\sigma_--\tfrac{1}{2}
\left\{\hat\sigma_- \hat\sigma_+ ,\hat \rho\right\}
\right) 
\nonumber \\
  + \gamma \bar m &
\left(\hat \sigma_+ \hat \rho \hat\sigma_--\tfrac{1}{2}
\left\{\hat\sigma_- \hat\sigma_+ ,\hat \rho\right\}
\right),
\end{align}
and its eigenvectors are the following four elements
\begin{align}
\check{\rho}_{0}=\hat{\mathbb I},\quad 
\check\rho_z=\tfrac{1}{2}\hat\sigma_z -\tfrac{2\bar m-1}{2}\hat{\mathbb I}, \quad
\check{\rho}_\pm=\hat{\sigma}_\pm.
\end{align}
The eigenvectors of $\cL_{\rm S}$ and $\cL_{\rm S}^\dagger$ belong to the space of $2 \times 2$ matrices $\mathcal{M}_2(\mathbb{C})$ and form a biorthogonal system as it can be seen that
they fulfill the relation
\begin{align}
\langle\check\rho_i,\hat\rho_j\rangle=\Tr{\check\rho_i^\dagger \hat{\rho}_j }=\delta_{i,j}
\label{}
\end{align}
with the Kronecker delta for all $i,j \in \{0,z,+,-\}$.

\subsection{Eigensystem of $\cL_{\rm HO}$}

The master equation describing a Markovian damped harmonic oscillator can be written
in the following form
\begin{align}
\cL_{\rm HO} \hat{\rho}=
\gamma (\bar n+1) &\left(\hat{a} \hat{\rho} \hat{a}^\dagger -\tfrac{1}{2}
\{\hat{a}^\dagger \hat{a}, \hat{\rho}\}\right) \nonumber \\
+ \gamma \bar n &\left(\hat{a}^\dagger \hat{\rho} \hat{a
}-\tfrac{1}{2}\{ \hat{a} \hat{a}^\dagger, \hat{\rho}\}\right).
\end{align}
The eigenvalue equations of $\cL_{\rm HO}$ and $\cL_{\rm HO}^\dagger$ read 
\begin{align}
  \cL_{\rm HO}\hat\rho_{n,k}=\lambda_{n,k}\hat\rho_{n,k},\quad
  \cL_{\rm HO}^\dagger\check\rho_{n,k}=\lambda_{n,k}^\ast\check\rho_{n,k}.
\label{}
\end{align}
The first equation is solved by the following right eigenvectors  (see Ref. \cite{BE})
\begin{equation}
\hat{\rho}_{n,k}=
  \hat{a}^{\dagger \frac{|k|+k}{2}}\tfrac{(-1)^n}{(\bar n +1)^{k+1}} :L^{(|k|)}_n
  \left(\tfrac{\hat{a}^\dagger \hat{a}}
  {\bar n +1}\right) e^{-\tfrac{\hat{a}^\dagger \hat{a}}{\bar n +1}}:\hat a^{\dagger\frac{|k|-k}{2}} 
\end{equation}
where $L^{(k)}_n$ is a generalized Laguerre polynomial and\\ $:f(\hat{a}^\dagger \hat{a}):$ is the normal ordering of $f$, a function of the number operator
$\hat{a}^\dagger \hat{a}$. The eigenvalues are 
\begin{eqnarray}
  \lambda_{n,k}=-\gamma \left(n+\frac{|k|}{2}\right), 
  \quad n\in \mathbb{N}_0,\quad k\in \mathbb{Z}. 
\end{eqnarray}
$\cL_{\rm HO}^\dagger$ is a representation of the adjoint of $\cL_{\rm HO}$ with the help of the trace as a linear functional
and can be found to be 
\begin{align}
\cL_{\rm HO}^\dagger \hat{\rho}=
\gamma (\bar n+1) &\left(\hat{a}^\dagger \hat{\rho} \hat{a} -\tfrac{1}{2}
\{\hat{a}^\dagger \hat{a}, \hat{\rho}\}\right) \nonumber \\
+ \gamma \bar n &\left(\hat{a} \hat{\rho} \hat{a}^\dagger
-\tfrac{1}{2}\{ \hat{a} \hat{a}^\dagger ,\hat{\rho}\}\right),
\end{align}
and its eigenvectors are given by 
\begin{equation}
\check{\rho}_{n,k}=
  \tfrac{(-\bar n)^n n!}{(\bar n +1)^n(n+k)!} 
 \hat{a}^{\frac{|k|+k}{2}} 
  :L^{(|k|)}_n\left(\tfrac{\hat{a}^\dagger \hat{a}}
 {\bar n }\right): \hat{a}^{\frac{|k|-k}{2}}. 
\end{equation}
It can be verified that they are orthogonal with respect to the inner product
\begin{align}
  \langle\check\rho_{n,k},\hat\rho_{n',k'}\rangle=
\Tr{\check{\rho}_{n,k}^\dagger\hat{\rho}_{n',k'}}=\delta_{n,n'}\delta_{k,k'},
  \label{}
\end{align}
and this is consistent with 
\begin{equation}
\langle\cL_{\rm HO}^\dagger \check\rho_{n,k},\hat\rho_{n,k}\rangle=\langle\check\rho_{n,k},\cL_{\rm HO}\hat\rho_{n,k}\rangle=\lambda_{n,k}. 
\end{equation}

\subsection{Comments on the eigensystem of $\cL_{\rm H0}$}

{\it Banach space structure.-}
In the case of the damped harmonic oscillator one has to deal with an 
infinite dimensional Hilbert space. 
The Liouville operator is a map on the set of trace class operators 
which do not form a Hilbert space, but do form a Banach space. 
Geometrical problems are inherent
in the concept of a Banach space and therefore a slightly different notion of adjoint operators is needed. 

Let $E,F$ be Banach spaces and if $T:E\rightarrow F$ is a bounded linear operator 
then we define its adjoint by
\begin{equation}
T': F' \rightarrow E',\, T'(\varphi)\rightarrow \varphi\circ T,\, \varphi \in F'
\end{equation}
where $T'$ is mapping bounded linear functionals on $F$ to bounded linear functionals on $E$ and $E'$($F'$) is the dual of $E$($F$).
The set of trace class operators  on a Hilbert space $\mathcal{H}$ is defined as
\begin{align}
  \cB_1(\cH):=\{\hat x \in \cB(\cH):\, \Tr{\sqrt{\hat x^\dagger \hat x}} <\infty\}.
\end{align}  
If $\cL: \mathcal{B}_1(\mathcal{H}) \rightarrow \mathcal{B}_1(\mathcal{H})$ is a bounded linear operator then the adjoint $\cL'$ 
is defined on the dual $\mathcal{B}_1(\mathcal{H})'$. Due to the fact that
the set of trace class operator form a two-sided ideal in $\mathcal{B}(\mathcal{H})$ (see Theorem VI.19 in Ref. \cite{Reed}) we have an isometric isomorphism
\begin{equation}
 \mathcal{B}_1(\mathcal{H})'\cong\mathcal{B}(\mathcal{H}).
\end{equation}
Thus, we may translate the adjoint $\cL'$ to an operator $\mathcal{L^\dagger}:\mathcal{B}(\mathcal{H}) 
\rightarrow \mathcal{B}(\mathcal{H})$ satisfying
\begin{equation}
 \mathrm{Tr}\{(\mathcal{L}^\dagger \hat{y})^\dagger \hat{\rho}\}=
 \mathrm{Tr}\{\hat{y}^\dagger (\mathcal{L} \hat{\rho})\}
\end{equation}
for all $\hat{y} \in \mathcal{B}(\mathcal{H})$ and $\hat{\rho} \in \mathcal{B}_1(\mathcal{H})$.

We can already conclude that the eigenvectors of $\cL_{\rm HO}'$ can be understood as the linear functional
$\varphi$ for a fixed $\check \rho_{n,k}$ as
\begin{equation}
  \varphi[\check{\rho}_{n,k}](\hat \rho)=
  \mathrm{Tr}\{\hat \varrho _{n,k}^\dagger \hat \rho\},\,\, \forall \hat\rho \in 
  \mathrm{Dom}(\cL_{\rm HO}).
\end{equation}

{\it Domain.-}
When dealing with infinite dimensional systems,  
the domain of the operators is a non-trivial concept that has to be handled carefully
as we are going to elucidate in the following lines.
The Hilbert space of the harmonic oscillator is the symmetric Fock space 
$\mathcal{H}=\Gamma_s(\mathbb{C})$. 
Every element can be expressed in terms of 
the orthonormal basis $\{\ket{n}\}_{n \in \mathbb{N}}$ as
\begin{eqnarray}
 x=\sum^\infty_{n=0} x_n \ket{n} \in \mathcal{H},\,\,\,  \sum^\infty_{n=0} |x_n|^2 < \infty.
\end{eqnarray}
The creation  and the annihilation operators ($\hat{a}^\dagger,\,\hat{a}$) have the domain
\begin{equation}
 \mathrm{Dom}(\hat{a}^\dagger)=\mathrm{Dom}(\hat{a})=\Big\{x:\, \sum^\infty_{n=1} n|x_n|^2 < \infty \Big\}
\end{equation}
which is dense in $\mathcal{H}$. The number operator $\hat{a}^\dagger \hat{a}$ has a different and more
restricted domain given by
\begin{equation}
\mathrm{Dom}(\hat{a}^\dagger \hat{a})=\Big\{x:\, \sum^\infty_{n=1} n^2|x_n|^2 < \infty \Big\}. 
\end{equation}
In order to exemplify this difference, let us define the following three vectors in $\mathcal{H}$:
\begin{eqnarray}
 x_1&=&\sum^\infty_{n=0} \frac{1}{n} \ket{n}, \,\,\,\, x_2=\sum^\infty_{n=0} \frac{1}{n^{3/2}} \ket{n}, \nonumber \\
 x_3&=&\sum^\infty_{n=0} \frac{1}{n^2} \ket{n}.
\end{eqnarray}
These vectors have the following properties
\begin{eqnarray}
 x_1 \notin \mathrm{Dom}(\hat{a}),\,\, x_1 \notin \mathrm{Dom}(\hat{a}^\dagger \hat{a}), \nonumber \\
 x_2 \in \mathrm{Dom}(\hat{a}),\,\, x_2 \notin \mathrm{Dom}(\hat{a}^\dagger \hat{a}), \nonumber \\
 x_3 \in \mathrm{Dom}(\hat{a}),\,\, x_3 \in \mathrm{Dom}(\hat{a}^\dagger \hat{a}).
\end{eqnarray}
This is because even when $x_1$ is a normalized vector, 
$\hat{a} x_1\notin\cH$ and $\hat{a}^\dagger \hat{a} x_1\notin\cH$
are not normalized anymore and therefore they are not in $\cH$.
In the case of $x_2$, it can be noted that $\hat{a} x_2\in \cH$ but $\hat{a}^\dagger \hat{a} x_2\notin\cH$.
With this observation we can conclude that 
$\mathrm{Dom}(\hat{a}^\dagger \hat{a}) \subset \mathrm{Dom}(\hat{a}) \subset \mathcal{H}$.

In the case of the Liouville operator and for the sake of simplicity, 
let us consider the case of zero a temperature bath ($\bar n=0$). In this case
the Liouvillian reads
\begin{equation}
 \mathcal{L}_0 \hat{\rho}=\hat{a} \hat{\rho} \hat{a}^\dagger-\frac{1}{2} \left(\hat{a}^\dagger \hat{a} \hat{\rho}+ \hat{\rho}\hat{a}^\dagger \hat{a}  \right).
 \label{exampleofL}
\end{equation}
In order to show that not all elements of $\mathcal{B}_1(\mathcal{H})$ are in the domain of $\cL_0$ , we consider the following density matrix
\begin{eqnarray}
 &&\hat{\rho}=\frac{6}{\pi^2}\sum^\infty_{n,m=1} \frac{1}{nm} \ket{n}\bra{m}, \nonumber \\
\end{eqnarray}
The normalization condition is fulfilled as 
$\sum^\infty_{n=1}{n^{-2}}=\pi^2/6$. Applying $\cL_0 $ one gets
\begin{eqnarray}
&&\hat{\rho}_\star=\mathcal{L}_0 \hat{\rho}=\frac{6}{\pi^2}\sum^\infty_{n=0}\frac{1}{\sqrt{n+1}} (\ket{n}\bra{0}+ \ket{0}\bra{n})+ \nonumber \\
&&+\frac{6}{\pi^2}\sum^\infty_{n,m=1} \left(\frac{1}{\sqrt{(n+1)(m+1)}}-\frac{n+m}{2nm}\right)\ket{n}\bra{m}. \nonumber
\end{eqnarray}
Now, let us remember that the normed space of Hilbert-Schmidt operators is a subset
of all bounded linear operators defined on a Hilbert space $\mathcal{H}$ and at the same time
it contains the set trace class operators, i.e.
\begin{equation}
 \mathcal{B}_1(\mathcal{H}) \subseteq \mathcal{B}_2(\mathcal{H}) \subseteq\mathcal{B}(\mathcal{H}).
\end{equation}
In order to determine if $\hat{\rho}_\star$ is a Hilbert-Schmidt operator we evaluate
\begin{eqnarray}
 \mathrm{Tr} \{\hat{\rho}^\dagger_\star \hat{\rho}_\star\}&=&\sum^\infty_{n,m=1}\left(\frac{1}{\sqrt{(n+1)(m+1)}}-\frac{n+m}{2nm}\right)^2 \nonumber \\
 &+&2\sum^\infty_{n=2}\frac{1}{n}+1
\end{eqnarray}
which is clearly diverging. Therefore $\hat{\rho}_\star \notin \mathcal{B}_2(\mathcal{H})$ 
implying that 
$\hat{\rho}_\star \notin \mathcal{B}_1(\mathcal{H})$. So $\mathcal{L}_0$ in Eq. \eqref{exampleofL}
cannot be defined on the whole space $\mathcal{B}_1(\mathcal{H})$, but only on its domain $\mathrm{Dom}(\mathcal{L}_0)$. The properties of $\mathrm{Dom}(\cL_{\rm HO})$ are not 
studied here, but one can at least state that 
\begin{eqnarray}
 \mathrm{Dom}(\cL_{\rm HO}) \subset \mathcal{B}_1(\mathcal{H}), \nonumber 
 \end{eqnarray}
 which implies that the bounded linear functionals on $\mathrm{Dom}(\cL_{\rm HO})$ regarding to the norm of $\mathcal{B}_1(\mathcal{H})$ are more numerous than the
 bounded linear functionals on $ \mathcal{B}_1(\mathcal{H})$.

{\it Linear independence and completeness.-}
The eigenvectors $\hat{\rho}_{n,k}$ due to the normal ordered 
form $:L^{(k)}_n(z \hat{a}^\dagger \hat{a})\ee^{-z\hat{a}^\dagger \hat{a}}:$ with $z\in(0,1]$ are trace class operators. Nonetheless, $(\hat{a}^\dagger \hat{a})^m \hat{\rho}_{n,k}$
is also a trace class operator for any finite value of $m$.
All the eigenvalues of $\cL_{\rm HO}$ are distinct which implies that the eigenvectors $\hat \rho_{n,k}$ are linearly independent. We have a countable infinite set of eigenvectors 
whose linear independence property 
is equivalent to show that all possible finite selection of eigenvectors are
linearly independent. 
In order to show this,
let $\{\hat{\rho}_{i(1)},\hat{\rho}_{i(2)},...,\hat{\rho}_{i(k)}\}$ be a finite set of 
eigenvectors with corresponding eigenvalues $\{\lambda_{i(1)},\lambda_{i(2)},...,\lambda_{i(k)}\}$ and let $c_1,c_2,...,c_k$ be scalars such that
\begin{equation}
 c_1\hat{\rho}_{i(1)}+c_2\hat{\rho}_{i(2)}+...+c_k\hat{\rho}_{i(k)}=0.
\end{equation}
$\cL_{\rm HO}$ has non-degenerate eigenvalues and applying 
$N(\cL_{\rm HO})=(\cL_{\rm HO}-\lambda_{i(2)})(\cL_{\rm HO}-\lambda_{i(3)})...
(\cL_{\rm HO}-\lambda_{i(k)})$ to the above equation results that all terms except the first are annihilated, resulting
\begin{equation}
  c_1N(\lambda_{i(1)})\hat{\rho}_{i(1)}=0,\,\,N(\lambda^A_{i(1)})\neq 0,
\end{equation}
and since $\hat{\rho}_{i(1)}\neq 0$ it follows that $c_1=0$. Similarly, $c_2=...=c_k=0$, which proves the linear independence of 
$\{\hat{\rho}_{i(1)},\hat{\rho}_{i(2)},...,\hat{\rho}_{i(k)}\}$. This proof can be applied to all possible finite sets of eigenvectors. The eigenvectors
are not orthogonal in regard to the Hilbert-Schmidt inner product. The set of eigenvectors has maximal possible cardinality, so $\{\hat \rho_{n,k}\}_{n \in \mathbb{N}, k \in \mathbb{Z}}$
is a basis.

{\it Spectrum.-}
Finite rank operators are dense in the set of trace class operators (see the Corollary after Theorem VI.21 in Ref. \cite{Reed}). The
eigenvectors $\hat{\rho}_{n,k}$ are finite rank operators and they form a basis which implies that 
$\cL_{\rm HO}$ is a closed densely defined operator. The point spectrum of  $\cL^\dagger_{\rm HO}$ coincides with the point spectrum of $\cL_{\rm HO}$ and this means that 
$\cL_{\rm HO}$ has no residual spectrum (see Proposition $1.12$ in Chap. IV of Ref. \cite{Engel}).
Now we know the point spectrum and the residual spectrum, but we have not yet discussed 
all possible approximate eigenvalues. $\lambda$ is an approximate eigenvalue if there exists a sequence 
$\{\hat x_n\}_{n \in \mathbb{N}} \subset \mathrm{Dom}(\cL_{\rm HO})$ such that $\|\hat x_n\|_1=1$ and $\lim_{n \to \infty}\|\cL_{\rm HO} \hat x_n-\lambda \hat x_n\|_1=0$. The latter equation can be
expanded in regard to the basis $\{\hat \rho_{n,k}\}_{n \in \mathbb{N}, k \in \mathbb{Z}}$ and yields
\begin{equation}
\lim_{n \to \infty} \|\sum_{n,k} \left(\lambda_{n,k}-\lambda\right) 
\mathrm{Tr}\{\hat \varrho_{n,k} \hat x_n\} \hat \rho_{n,k} \|_1=0, \nonumber
\end{equation}
which is fulfilled only when $\lambda_{n,k}=\lambda$, meaning that $\cL_{\rm HO}$ has only a point spectrum.

\section{Steady state of a damped harmonic oscillator}
\label{Appendix}

Let us consider the model of a damped harmonic oscillator given by the master equation
\begin{eqnarray}
\label{damp}
 \frac{d\hat{\rho}}{dt}=&-&\frac{\gamma_1}{2} \big(\hat{a}^\dagger \hat{a} \hat{\rho}-2 \hat{a} \hat{\rho} \hat{a}^\dagger+ \hat{\rho}
\hat{a}^\dagger \hat{a}\big) \nonumber \\
&-& \frac{\gamma_2}{2} \big( \hat{a} \hat{a}^\dagger \hat{\rho}-2 \hat{a}^\dagger \hat{\rho} \hat{a}+ \hat{\rho}
\hat{a} \hat{a}^\dagger\big)
\end{eqnarray}
where $\hat{a}$ ($\hat{a}^\dagger$) is the annihilation (creation) operator of the harmonic oscillator. 
$\gamma_1$ and $\gamma_2$ have the dimension of a frequency. To ensure 
that Eq. \eqref{damp} describes the evolution of a density matrix 
we require the condition $\gamma_2 < \gamma_1$.

To investigate in more detail the steady state of the damping, we consider the density matrix of the field in number state representation $|n\rangle$ ($n\in {\mathbb N}_0$)
\begin{equation}
 \hat{\rho}_{\rm st}=
 \sum^{\infty}_{n,m=0} \rho_{n,m} \ket{n}\bra{m}.
\end{equation}
We set the left-hand side of Eq. \eqref{damp} to zero, and taking matrix elements of the density matrix we get
\begin{eqnarray}
\label{sdamp}
0=&-&\frac{\gamma_1}{2} (m+n)  \rho_{n,m} + \gamma_1 \sqrt{(n+1)(m+1)}  \rho_{n+1,m+1} \nonumber \\
&-&\frac{\gamma_2}{2} (m+1+n+1)  \rho_{n,m} + \gamma_2 \sqrt{nm}  \rho_{n-1,m-1}. \nonumber \\
\end{eqnarray}
The matrix elements are coupled via a second order recurrence relation to their nearest neighbours in each of the diagonals of the density matrix. 

First, we consider the 
main diagonal ($m=n$) and Eq. \eqref{sdamp} simplifies to
\begin{equation}
\rho_{n+1,n+1}=
\frac{\gamma_1 n + 
\gamma_2 (n+1)}{\gamma_1 (n+1)} 
\rho_{n,n}  
- \frac{\gamma_2 n}{\gamma_1(n+1)}  \rho_{n-1,n-1}.
\end{equation}
The solution can be determined by induction, thus yielding
\begin{equation}
\label{diagsur}
\rho_{n,n}=\left(\gamma_2/\gamma_1\right)^n \rho_{0,0},\quad n\geq0, 
\end{equation}
The normalization condition $\mathrm{Tr}\{\hat{\rho}\}=1$ results in
\begin{equation}
\rho_{0,0} \sum^{\infty}_{n=0}\left(\frac{\gamma_2}{\gamma_1}\right)^n=1.  
\end{equation}
With the geometric series
\begin{equation}
\sum^{\infty}_{n=0}\left(\frac{\gamma_2}{\gamma_1}\right)^n=\frac{1}{1-\gamma_2/\gamma_1},
\end{equation}
one finds that $\rho_{0,0}=(\gamma_1-\gamma_2)/\gamma_1$.

Now, we consider only the upper diagonals $m>n$ due to the self-adjoint property 
of the density matrix $\hat{\rho}_s$.  Defining $l=m-n\geq1$ 
and $\gamma_2=\epsilon \gamma_1$ with
$\epsilon \in [0,1)$ Eq. \eqref{sdamp} takes the form
\begin{align}
\label{sdampoff}
\rho_{n,n+l}=&
r_{l,n}^{\epsilon}\rho_{0,l},
\nonumber\\
r_{l,n+1}^{\epsilon}=&
\left(f^n_l +\epsilon g^n_l\right)
r_{l,n}^{\epsilon} 
-\epsilon h^n_l  r_{l,n-1}^{\epsilon}
\end{align}
where  we introduced the following abbreviations
\begin{eqnarray}
\label{fgh}
f^n_l&=&\left(n+l/2\right)/\sqrt{(n+1)(n+l+1)},\nonumber \\
g^n_l&=&\left(n+1+l/2\right)/\sqrt{(n+1)(n+l+1)},\nonumber \\
h^n_l&=&\sqrt{n(n+l)}/\sqrt{(n+1)(n+l+1)},
\end{eqnarray}
which are all positive and fulfill the following inequalities
\begin{align}
  f^n_l-h^n_l>0,\quad g^n_l>1, \quad \forall\,\, n \geq0 ,\,l\geq 1.
  \label{fghprop}
\end{align}

For the first few terms it can be shown that they have solutions of the form
\begin{equation}
\label{offform}
 r_{l,n}^{\epsilon}= \sum^n_{i=0} a^{n,l}_i \epsilon^i,
\end{equation}
with coefficients obeying the properties
\begin{equation}
\label{offcond}
a^{n,l}_i > 0,\,\, a^{n-1,l}_i<a^{n,l}_{i+1},\,\, \forall n\geq i \geq0.
\end{equation}
We apply a proof by induction to show that $r_{l,n+1}^{\epsilon}$ takes the form of Eq. \eqref{offform} with the conditions of Eq. \eqref{offcond}.
First we use Eq. \eqref{sdampoff} which results in
\begin{equation}
r_{l,n+1}^{\epsilon}=\sum^n_{i=0} f^n_l a^{n,l}_i \epsilon^i + \sum^{n}_{i=0} g^n_l a^{n,l}_i \epsilon^{i+1}-\sum^{n-1}_{i=0} h^n_l a^{n-1,l}_i \epsilon^{i+1},   
\end{equation}
and translates to equations for the coefficients:
\begin{eqnarray}
 a^{n+1,l}_0&=&f^n_l a^{n,l}_0, \\
 a^{n+1,l}_i&=&f^n_l a^{n,l}_i + g^n_l a^{n,l}_{i-1}- h^n_l a^{n-1,l}_{i-1}, i\geq 1. \nonumber
\end{eqnarray}
Now, applying the conditions of Eqs. \eqref{fghprop}, and \eqref{offcond} we find that
\begin{equation}
 a^{n+1,l}_i > 0,\,\,  a^{n,l}_i<a^{n+1,l}_{i+1}
\end{equation}
and it follows immediately that every $r_{l,n}^{\epsilon}$ can be cast in the form of Eq. \eqref{offform}. 

As all coefficients are positive, the sum is always larger than the zeroth order
coefficient
\begin{align}
  r_{l,n}^{0}=\prod_{j=0}^{n-1}f^n_j=
  \frac{\Gamma\left(n+\tfrac{l}{2}\right)}{\Gamma\left(\tfrac{l}{2}\right)}
  \sqrt{\frac{l!}{n!(n+l)!}},
  \label{a0def}
\end{align}
which we have expressed in terms of the gamma function $\Gamma(z)$, and multiplied
by $\rho_{0,l}$ 
solves the first order recurrence relation obtained from the Eq. 
\eqref{sdamp} in the case when $\epsilon=0$ which implies $\gamma_2=0$.

These considerations allow us to set a lower bound to the following sum:
\begin{equation}
  |\sum^{\infty}_{n=0}\rho_{n,n+l}|
  \geq |\rho_{0,l}| \sum^{\infty}_{n=0} r_{l,n}^{0}
  \geq |\rho_{0,l}| \sum^{\infty}_{n=0} r_{1,n}^{0},
\end{equation}
as the relation $r_{l,n}^{0}\geq r_{1,n}^{0}$ 
can be verified by noting that each term in the product of Eq. \eqref{a0def}
has the property $f^n_l\geq f^n_1$ which can be checked from Eq. \eqref{fgh}.
Stirling's formula in an inequality form \cite{Nanjundiah,Kazarinoff}
\begin{equation}
\sqrt{2 \pi n} \left(\frac{n}{e}\right)^n < n!< \sqrt{2 \pi n} \left(\frac{n}{e}\right)^n e^{\frac{1}{12n}}
\end{equation}
and the relation $\Gamma(n+1/2)=\sqrt\pi(2n)!/(4^n n!)$ yield 
\begin{equation}
r_{1,n}^{0}
>\frac{e^{-\frac{1}{6n}}}{\sqrt{\pi n(n+1)}}
>\frac{1}{\sqrt{\pi e^{1/3}}(n+1)}. \nonumber
\end{equation}
Thus, we have found that all the solutions to Eq. \eqref{sdamp} 
for $l\neq 0$ fulfill the inequality
\begin{equation}
\label{finalineq}
|\sum^{\infty}_{n=0} \rho_{n,n+l}|>|\rho_{0,l}|\left(1+\sum^{\infty}_{n=1}\frac{1}{\sqrt{\pi \ee^{1/3}}(n+1)} \right).
\end{equation}
The right-hand side of the above inequality is divergent for $|\rho_{0,l}|>0$ because it contains a harmonic series. 

In the next step we show that Eq. \eqref{finalineq} is upper bounded by $1$ due to some general facts of operator theory. 
Let us consider an arbitrary separable Hilbert space $\mathcal{H}$ with an orthonormal basis $(e_i)$. 
The set of Hilbert-Schmidt operators $\mathcal{B}_2(\mathcal{H})$ with the Hilbert-Schmidt inner product
\begin{equation}
 \langle \hat{X},\hat{Y} \rangle=\sum_i\langle \hat{X}e_i,\hat{Y}e_i \rangle=\mathrm{Tr}\{\hat{X}^\dagger \hat{Y}\},\,\, \forall \hat{X},\hat{Y} \in \mathcal{B}_2(\mathcal{H})
\end{equation}
is a Hilbert space and it is also a two-sided ideal in $\mathcal{B}(\mathcal{H})$(see Theorem VI.22 in Ref. \cite{Reed}):
\begin{equation}
\label{two-sided}
 ||\hat{A} \hat{X}||_2, ||\hat{X} \hat{A}||_2<\infty,\,\hat{X} \in \mathcal{B}_2(\mathcal{H}),\,\hat{A} \in \mathcal{B}(\mathcal{H}). 
\end{equation}
In our case the Hilbert space $\mathcal{H}=\Gamma_s(\mathbb{C})$ and any density matrix $\hat{\rho} \in \mathcal{B}_2(\mathcal{H})$ ($|| \hat{\rho} ||_2 \leq 1$) and
$\hat{\rho}^{1/2} \in \mathcal{B}_2(\mathcal{H})$ ($|| \hat{\rho}^{1/2} ||_2 = 1$). 

Now let us consider the right shift operator
\begin{eqnarray}
 &&\hat{S}_R \ket{n}=\ket{n+1}, \, n\geq0, \nonumber \\
 &&\hat{S}_R \in \mathcal{B}(\mathcal{H}).
\end{eqnarray}
Any $l$th power of the right shift operator $\hat{S}^l_R$ is also a bounded operator. Then
\begin{equation}
| \mathrm{Tr} (\hat{S}^l_R \hat{\rho}_s )|=|\sum^\infty_{n=0} \rho_{n,n+l}|
\end{equation}
and $\hat{S}^l_R \hat{\rho}_s \in \mathcal{B}_2(\mathcal{H})$. The Cauchy-Schwarz-Bunyakovski inequality for the Hilbert-Schmidt inner product gives:
\begin{equation}
 |\mathrm{Tr} (\hat{X}^\dagger Y)|=|\langle \hat{X},\hat{Y} \rangle | \leq ||X||_2 ||Y||_2, \,\, \forall \hat{X},\hat{Y} \in \mathcal{B}_2(\mathcal{H}).
\end{equation}
We observe that for the Hilbert-Schmidt operators $\hat{X}=\hat{\rho}^{1/2}_{\rm st}$ and $\hat{Y}=\hat{S}^l_R \hat{\rho}^{1/2}_{\rm st}$ 
\begin{equation}
 | \mathrm{Tr} (\hat{S}^l_R \hat{\rho}_s )| \leq ||\hat{\rho}^{1/2}_{\rm st}||_2 ||\hat{S}^l_R \hat{\rho}^{1/2}_{\rm st}||_2=1,
\end{equation}
where we used the relation $\left(\hat{S}^l_R\right)^\dagger\hat{S}^l_R=\hat{I}$, the identity operator on $\mathcal{H}$. Therefore, Eq. \eqref{finalineq} obviously leads to
\begin{equation}
 1 \geq |\sum^\infty_{n=0} \rho_{n,n+l}| > |\rho_{0,l}|+\frac{1}{\sqrt{\pi \ee^{1/3}}} |\rho_{0,l}| \sum^\infty_{n=1} \frac{1}{n+1}. 
\end{equation}
which is true if and only if $\rho_{0,l}=0$. If $\rho_{0,l}=0$, then all the upper diagonal elements of $\hat{\rho}_{\rm st}$ are zero. 
It is immediate from the self-adjoint property of $\hat{\rho}_{\rm st}$ that
all lower diagonal elements are also equal to zero. Finally, we can conclude that $\hat{\rho}_{\rm st}$ has only non-zero elements in the diagonal which are given in Eq. \eqref{diagsur}.

\end{document}